\documentclass[journal,10pt,twocolumn,twoside]{IEEEtran}
\ifCLASSINFOpdf
  \usepackage[pdftex]{graphicx}
  \DeclareGraphicsExtensions{.pdf,.jpeg,.png}
\else
  \usepackage[dvips]{graphicx}
  \DeclareGraphicsExtensions{.eps}
\fi
\usepackage[cmex10]{amsmath}
\usepackage{amsfonts}
\usepackage{array}

\usepackage[tight,footnotesize]{subfigure}
\usepackage{url}

\usepackage{bm}
\usepackage{psfrag}
\usepackage{verbatim}
\newtheorem{prop}{Proposition}
\newtheorem{lemma}{Lemma}

\providecommand{\abs}[1]{\left\lvert#1\right\rvert}

\DeclareMathOperator{\E}{\mathbf{E}}

\DeclareMathOperator*{\argmin}{\arg\,\min}

\newcommand{\mbI}{\mathbf{I}}

\newcommand{\mbx}{\mathbf{x}}

\newcommand{\mbmu}{\boldsymbol{\mu}}

\newcommand{\mby}{\mathbf{y}}

\newcommand{\mbp}{\mathbf{p}}

\newcommand{\mbz}{\mathbf{z}}
\newcommand{\mbY}{\mathbf{Y}}
\newcommand{\mblambda}{\boldsymbol{\lambda}}
\newcommand{\mbtheta}{\boldsymbol{\theta}}
\newcommand{\mbn}{\mathbf{n}}
\newcommand{\mbsigma}{\boldsymbol{\sigma}}
\newcommand{\ones}{\mathbf{1}}
\newcommand{\mbkappa}{\boldsymbol{\kappa}}

\newcommand{\betat}{\widetilde{\beta}}

\newcommand{\Jt}{\widetilde{J}}

\newcommand{\thetah}{\hat{\theta}}

\newcommand{\lambdab}{\overline{\lambda}}
\newcommand{\mblambdab}{\overline{\boldsymbol{\lambda}}}
\newcommand{\hb}{\overline{h}}

\begin{document}
%
\title{Multistage Adaptive Estimation of Sparse Signals}
%
%
%

\author{Dennis~Wei~and~Alfred~O.~Hero,~III
\thanks{Copyright \copyright\ 2013 IEEE. Personal use of this material is permitted. However, permission to use this material for any other purposes must be obtained from the IEEE by sending a request to pubs-permissions@ieee.org.

This work was partially supported by Army Research Office grant W911NF-11-1-0391.  

The authors are with the Department
of Electrical Engineering and Computer Science, University of Michigan, Ann Arbor,
MI 48109 USA (e-mail: dlwei@eecs.umich.edu, hero@eecs.umich.edu).}}
\maketitle

\begin{abstract}
This paper considers sequential adaptive estimation of sparse signals under a constraint on the total sensing effort.  The advantage of adaptivity in this context is the ability to focus more resources on regions of space where signal components exist, thereby improving performance.  A dynamic programming formulation is derived for the allocation of sensing effort to minimize the expected estimation loss.  Based on the method of open-loop feedback control, allocation policies are then developed for a variety of loss functions.  The policies are optimal in the two-stage case, generalizing an optimal two-stage policy proposed by Bashan \textit{et al.}, and improve monotonically thereafter with the number of stages.  Numerical simulations show gains up to several dB as compared to recently proposed adaptive methods, and dramatic gains compared to non-adaptive estimation.  An application to radar imaging is also presented.
\end{abstract}

\begin{IEEEkeywords}
Adaptive sensing, adaptive sampling, resource allocation, sparse signals, dynamic programming.
\end{IEEEkeywords}

%
\IEEEpeerreviewmaketitle


\section{Introduction}
\label{sec:intro}
%
%
%
%

Adaptive sensing and inference have been gaining interest in recent years in signal processing and related fields.  Potentially substantial gains in performance can be achieved when observations are made sequentially and adaptively, making use of information derived from previous observations.  This work focuses on sparse signals, i.e., signals that occupy a small number of dimensions in an ambient space.  It is now well-known that compressed sensing offers an efficient non-adaptive strategy for acquiring sparse signals, relying on a relatively small number of observations that are incoherent with the basis in which the signal is sparse (see e.g.~\cite{candesrombergtao2006,donoho2006}).  However, when noise is present and sensing resources are limited, incoherent observations may not be the most efficient since a large fraction of the resources are allocated to dimensions where the signal is absent.  Alternatively, by allocating resources according to estimates of the signal support obtained from past observations, better signal-to-noise ratios (SNR) are possible.  Applications in which adaptive sensing of sparse signals can be readily utilized include surveillance using active radars \cite{bashan2008,bashan2011}, spectrum sensing in cognitive radio \cite{tajer2012,zhang2010}, and gene association and expression studies \cite{zehetmayer2008}. 

Existing methods for adaptive sensing of sparse signals can be roughly grouped around two classes of models.  In the first class, which is the focus of this paper, observations are restricted to single components in the basis that induces signal sparsity, while resources can be distributed arbitrarily over components and observation stages.  An optimal two-stage allocation policy was developed in \cite{bashan2008} for a cost function related to bounds on estimation and detection performance.  
Subsequent developments stemming from \cite{bashan2008} include a modification to handle non-uniform signal priors \cite{newstadt2010}, a simplification based on Lagrangian constraint relaxation \cite{hitchings2010}, and a multiscale approach that uses linear combinations in the first stage to reduce the number of measurements \cite{bashan2011}.  Based on a similar model but in a different direction, a method known as distilled sensing \cite{haupt2011} was proposed for signal support identification and was shown to be asymptotically reliable (as the ambient dimension increases) at SNR levels significantly lower than non-adaptive limits.  The distilled sensing idea was recently extended to a more general setting of sequential multiple hypothesis testing in \cite{malloy2011a}; in \cite{malloy2011b} it is shown that a sequential thresholding procedure comes within a small factor of the optimal sequential procedure in terms of the number of observations needed for asymptotically exact support recovery.  

In the second class of models, the observations can consist of arbitrary linear combinations, as in compressed sensing, but for the most part the resource budget is assumed to be discrete, measured in units of normalized observations (\cite{malloy2011a,malloy2011b} also assume a discrete budget).  In \cite{haupt2012}, the distilled sensing approach was extended to the compressed measurement setting.  In \cite{ji2008,castro2008}, a Bayesian signal model is adopted and each new observation is chosen to approximately maximize the information gain; \cite{castro2008} is computationally simpler but is most suited to signals with a single non-zero component, i.e., $1$-sparse signals.  Others have also taken the approach of decomposing the problem into subproblems involving $1$-sparse signals and then applying a form of bisection search \cite{aldroubi2009,iwen2010,indyk2011}; \cite{indyk2011} employs a more sophisticated search in which the rate of division accelerates, reducing the dependence of the number of observations on the dimension to doubly logarithmic instead of merely logarithmic.  The adaptive methods in \cite{iwen2010,indyk2011} were shown to require fewer measurements than the best non-adaptive method.  In \cite{aldroubi2009} and \cite{indyk2011} however, noise is either not considered or not fully taken into account.  
Somewhat different from the aforementioned works is \cite{malioutov2010}, which describes a compressed sensing method that is sequential in the sense that it terminates once the reconstruction error is determined to have fallen below a threshold, but the form of the measurements is not adapted during the process.

Adaptive sensing and resource allocation have also been applied to other classes of signals with more structure.  Tree-structured sparsity is considered in \cite{averbuch2012}, which proposes selective sampling of wavelet coefficients based on already sampled coefficients nearby and at coarser resolutions.  For two-dimensional piecewise-constant signals, a method that concentrates measurements near boundaries is presented and analyzed in \cite{castro2005,willett2004}.  Adaptive waveform amplitude design is investigated in \cite{rangarajan2007} for unstructured (i.e.~dense) parameter estimation in a linear Gaussian model under an average energy constraint. 

This paper addresses the problem of estimation and adaptive resource allocation under the first observation model in which components are measured directly.  
We extend the two-stage allocation policy in \cite{bashan2008} to an arbitrary number of stages, focusing on estimation error explicitly as contrasted with performance bounds in \cite{bashan2008}.  Our method is computationally tractable for a wide range of estimation loss functions satisfying a mild convexity condition, including such commonly used criteria as mean squared error (MSE) and mean absolute error (MAE).  The observation model in \cite{bashan2008,haupt2011} is also generalized by allowing the observation precision to depend on an arbitrary concave function of the sensing effort.  It is shown that the problem can be formulated as a dynamic program, a framework that facilitates the development of allocation policies.  An approximate dynamic programming solution is proposed based on open-loop feedback control (OLFC).  The performance of these OLFC policies improves monotonically with the number of stages, and in particular improves upon optimal two-stage policies including the one in \cite{bashan2008}.  Numerical simulations show error reductions up to $4.5$ dB relative to the optimal two-stage policy and dramatic reductions relative to non-adaptive sensing, approaching the oracle limit at high SNR.  The OLFC policies are also shown to outperform distilled sensing \cite{haupt2011} at all SNR and most significantly at higher SNR.  The advantages carry over to a radar imaging example that challenges some of the assumptions of our model.

The remainder of the paper proceeds as follows.  In Section \ref{sec:prob}, the signal and observation models are specified and a problem of resource-constrained sequential estimation is formulated and then recast as a dynamic program.  In Section \ref{sec:policy}, optimal and OLFC approaches to the problem are discussed and a family of OLFC policies is proposed.  Numerical simulations comparing our OLFC policies to other policies are presented in Section \ref{sec:perfEval}.  In Section \ref{sec:SARtanks}, an application to radar imaging is described.  Conclusions and future directions are given in Section \ref{sec:concl}.

\section{Problem formulation}
\label{sec:prob}

We consider signals $\mbtheta \in \mathbb{R}^{N}$ that are observed in the same basis in which they are sparse; the basis is taken to be the standard basis without loss of generality.
%
%
The signal support is represented by a set of indicators $I_{i}$, $i = 1, \ldots, N$, with $\theta_{i} = 0$ if $I_{i} = 0$. 
We use a probabilistic model in which $I_{i} = 1$ with prior probability $p_{i}(0)$, independently of the other indicators.  For $I_{i}=1$, the non-zero signal amplitudes $\theta_{i}$ are modelled as independent Gaussian random variables with prior means $\mu_{i}(0)$ and variances $\sigma_{i}^{2}(0)$.  As in \cite{bashan2008,bashan2011}, a non-informative uniform prior is assumed with $p_{i}(0) = p_{0}$, $\mu_{i}(0) = \mu_{0}$, and $\sigma_{i}^{2}(0) = \sigma_{0}^{2}$ for all $i$, although the theory developed below could also accommodate non-uniform priors.  

Observations are made in $T$ stages with non-negative effort levels $\lambda_{i}(t)$ that can vary with index $i$ and time $t = 0, \ldots, T-1$.   Depending on the application, the effort $\lambda_{i}(t)$ might represent observation time, number of samples, energy, cost, or computation.  It is assumed that the precision (inverse variance) of an observation varies with effort according to a non-decreasing function $h$ such that $h(0) = 0$, $h(\lambda) > 0$ for $\lambda > 0$, and normalized so that $h(1) = 1$.  For $\lambda_{i}(t-1) > 0$, the observation of the $i$th component at time $t$ takes the form 
\begin{equation}\label{eqn:obs}
y_{i}(t) = \theta_{i} + \frac{n_{i}(t)}{\sqrt{h(\lambda_{i}(t-1))}}, \quad i = 1, \ldots, N, \quad t = 1, \ldots, T,  
\end{equation}
where $n_{i}(t)$ represents i.i.d. zero-mean Gaussian noise with variance $\sigma^{2}$, whereas for $\lambda_{i}(t-1) = 0$ the observation is not taken.  Hence the number of observations per stage is at most $N$ but can be substantially lower if most of the $\lambda_{i}(t-1)$ are zero. The function $h$ is often linear, but nonlinear dependences can also arise.  For example, the sensing system may contain nonlinear components such as amplifiers, or the observations may result from integrating a continuous-time random process over an interval of length $\lambda_{i}(t-1)$ and the process exhibits short-term correlation.  
We restrict our attention to static signals so that the signal component $\theta_{i}$ in \eqref{eqn:obs} does not change with time.  For convenience, we use the vector notation $\mby(t) = \left[ y_{1}(t) \dots y_{N}(t) \right]^{T}$ (similarly for other indexed quantities) and denote by $\mbY(t) = \{ \mby(1), \ldots, \mby(t) \}$ the history of observations up to time $t$.

The task is to determine the distribution of sensing effort over components and time subject to a total budget constraint, 
\begin{equation}\label{eqn:totEnergy}
\sum_{t=0}^{T-1} \sum_{i=1}^{N} \lambda_{i}(t) = \Lambda_{0}.
\end{equation}
Under the normalization $\Lambda_{0} = N$, each component receives an average of one unit of effort over time.  In the case of single-stage non-adaptive estimation ($T = 1$) and a uniform prior, the most natural choice is to set $\lambda_{i}(0) = 1$ for all $i$.  Thus $\sigma^{2}$ can be regarded as the noise variance realized under a non-adaptive uniform allocation.  In multistage adaptive sensing, the allocation $\mblambda(t)$ at time $t$ can depend on the observations $\mbY(t)$ collected up to that point.  This information allows more resources to be focused on the region of signal support, thereby improving the SNR.  The mapping from $\mbY(t)$ to $\mblambda(t)$ is referred to as an effort allocation policy.  We restrict attention to deterministic policies in this work.  For notational brevity, we will not make the dependence of $\mblambda(t)$ on $\mbY(t)$ explicit.  

In this paper, we adopt the viewpoint that the nonzero signal components are of primary interest.  Thus our objective is to minimize the expected estimation loss over the signal support, 
%
\begin{equation}\label{eqn:expLossROI}
\E\left\{ \sum_{i=1}^{N} I_{i} L\left( \abs{\thetah_{i} - \theta_{i}} \right) \right\},
\end{equation}
where the estimates $\thetah_{i}$ are based on all observations up to time $T$, the loss function $L$ is non-decreasing, and the expectation is taken over $\mbI$, $\mbtheta$, and $\mbY(T)$.  
Under \eqref{eqn:expLossROI}, missed nonzero components are penalized directly through larger losses, while false alarms, i.e., zero-valued components mistaken as nonzero, are penalized indirectly because they divert resources away from the true signal support.

To relate the expected cost \eqref{eqn:expLossROI} to the effort allocation policy, we nest the expectations in the order $\mbY(T)$, $\mbI$, $\mbtheta$ (outer to inner) and expand to yield 
%
\begin{equation}\label{eqn:expLossROI2}
\E \left\{ \sum_{i=1}^{N} p_{i}(T) \E\left[ L\left( \abs{\thetah_{i} - \theta_{i}} \right) \mid I_{i} = 1, \mbY(T) \right] \right\},
\end{equation}
where we have defined $p_{i}(t) = \Pr(I_{i}=1 \mid \mbY(t))$.  We then make use of the following lemmas proved in Appendices \ref{app:probDist} and \ref{app:CME} respectively:
\begin{lemma}\label{lem:probDist}
The conditional amplitudes $\theta_{i} \mid I_{i} = 1, \mbY(t)$ remain independent Gaussian for all $t$ with means $\mu_{i}(t)$ and variances $\sigma_{i}^{2}(t)$.  Likewise, the conditional indicators $I_{i} \mid \mbY(t)$ remain independent Bernoulli for all $t$ with parameters $p_{i}(t)$.
\end{lemma}
\begin{lemma}\label{lem:CME}
If a random variable $\theta$ has a probability density $f(\theta)$ that is symmetric about $\mu$, i.e., $f(\mu-\theta) = f(\mu+\theta)$ for all $\theta$, and (weakly) unimodal, i.e., $f(\theta)$ is non-decreasing for $\theta < \mu$ and non-increasing for $\theta > \mu$, then $\thetah = \mu$ minimizes the expected loss $\E \left[ L\left( \bigl\lvert\thetah-\theta\bigr\rvert \right) \right]$ for any non-decreasing loss function $L$.
\end{lemma}
%

From Lemmas \ref{lem:probDist} and \ref{lem:CME} and the symmetry and unimodality of the Gaussian distribution, we conclude that the inner expectation in \eqref{eqn:expLossROI2} is minimized by choosing $\thetah_{i} = \mu_{i}(T)$ for $i = 1,\ldots,N$.  Then the minimum value of the inner expectation depends only on $\sigma_{i}^{2}(T)$ and \eqref{eqn:expLossROI2} can be expressed after a change of variables as
\begin{equation}\label{eqn:costFun}
2\E \left\{ \sum_{i=1}^{N} p_{i}(T) \int_{0}^{\infty} L\left( \sigma_{i}(T) \theta \right) \phi(\theta; 0,1) \, d\theta \right\},
\end{equation}
where $\phi(\theta; \mu, \sigma^{2})$ denotes the standard Gaussian probability density function with mean $\mu$ and variance $\sigma^{2}$. 
The final-stage variance $\sigma_{i}^{2}(T)$ depends in turn on the effort allocation according to the relation 
%
\begin{equation}\label{eqn:sigmaFinal}
\sigma_{i}^{2}(T) = \frac{\sigma^{2}}{\sigma^{2} / \sigma_{0}^{2} + \sum_{t=0}^{T-1} h(\lambda_{i}(t))},
\end{equation}
%
which follows from the proof of Lemma \ref{lem:probDist} in Appendix \ref{app:probDist}. 
%
%
In summary, the problem is to minimize the expected cost defined by \eqref{eqn:costFun} and \eqref{eqn:sigmaFinal} with respect to the effort allocation policy $\mblambda(0), \ldots, \mblambda(T-1)$, subject to the total effort constraint \eqref{eqn:totEnergy}.

In the case of the square loss $L(a) = a^{2}$, i.e., the mean squared error (MSE) criterion, the integral in \eqref{eqn:costFun} can be evaluated to yield $\sigma_{i}^{2}(T)$, thus reducing \eqref{eqn:costFun} to 
\begin{equation}\label{eqn:costFunMSE}
\sigma^{2} \E \left\{ \sum_{i=1}^{N} \frac{p_{i}(T)}{\sigma^{2}/\sigma_{0}^{2} + \sum_{t=0}^{T-1} h(\lambda_{i}(t))} \right\}.
\end{equation}
The cost function in \eqref{eqn:costFunMSE} is closely related to the cost function in \cite{bashan2008} although the motivations differ with the latter being related to Chernoff and Cram\'{e}r-Rao bounds on detection and estimation performance respectively.  The general form of the cost function in \cite{bashan2008} can be obtained from \eqref{eqn:costFunMSE} by replacing $p_{i}(T)$ with the weighted average $\nu p_{i}(T) + (1 - \nu) (1 - p_{i}(T))$ for $\nu \in [1/2, 1]$, letting $\sigma_{0}^{2} \to \infty$ so that $\sigma^{2}/\sigma_{0}^{2} \to 0$, and choosing $h$ to be the identity function.  
Given that the generalization of $p_{i}(T)$ to a weighted average is straightforward to accommodate, we keep $\nu = 1$ to simplify notation in the remainder of the paper.

\subsection{Formulation as a dynamic program}
\label{subsec:probDP}

The determination of an optimal effort allocation policy according to \eqref{eqn:costFun} and \eqref{eqn:sigmaFinal} 
can be formulated as a dynamic program.  Although the dynamic programming viewpoint does not offer significant simplifications, it does make available a well-developed set of approaches to the problem, some of which are considered in Section \ref{sec:policy}.  Further background in dynamic programming can be found in \cite{bertsekas2005}.

To formulate a sequential decision problem as a dynamic program, the cost function must be expressible as a sum of terms indexed by time $t$, where each term depends only on the current system state $\mbx(t)$ and the current control action, in our case the effort allocation $\mblambda(t)$ (each term may also depend on a random disturbance but this is not required here).  The cost function \eqref{eqn:costFun} can be recast in the required time-separable form by defining the state $\mbx(t)$ as $\mbx(t) = (\mbp(t), \mbmu(t), \mbsigma^{2}(t), \Lambda(t))$, 
%
%
where $\Lambda(t)$ represents the effort budget remaining at time $t$.  The state variables are initialized as $p_{i}(0) = p_{0}$, $\mu_{i}(0) = \mu_{0}$, $\sigma_{i}^{2}(0) = \sigma_{0}^{2}$, and $\Lambda(0) = \Lambda_{0}$, and evolve according to the following recursions derived in Appendix \ref{app:probDist}:
\begin{subequations}\label{eqn:stateEvol}
\begin{align}
p_{i}(t+1) &= \frac{p_{i}(t) \phi_{1}}{p_{i}(t) \phi_{1} + (1 - p_{i}(t)) \phi_{0}},\label{eqn:pEvol}\\
\mu_{i}(t+1) &= \frac{\sigma^{2} \mu_{i}(t) + h(\lambda_{i}(t)) \sigma_{i}^{2}(t) y_{i}(t+1)}{\sigma^{2} + h(\lambda_{i}(t)) \sigma_{i}^{2}(t)},\label{eqn:muEvol}\\
\sigma_{i}^{2}(t+1) &= \frac{\sigma^{2} \sigma_{i}^{2}(t)}{\sigma^{2} + h(\lambda_{i}(t)) \sigma_{i}^{2}(t)},\label{eqn:sigmaEvol}\\
\Lambda(t+1) &= \Lambda(t) - \sum_{i=1}^{N} \lambda_{i}(t),\label{eqn:LambdaEvol}
\end{align}
\end{subequations}
where 
\begin{align*}
\phi_{0} &= \phi(y_{i}(t+1); 0, \sigma^{2} / h(\lambda_{i}(t)) ),\\
\phi_{1} &= \phi(y_{i}(t+1); \mu_{i}(t), \sigma_{i}^{2}(t) + \sigma^{2} / h(\lambda_{i}(t)) ).
\end{align*}
%

Given the above state definition, 
we use \eqref{eqn:sigmaEvol} to rewrite the denominator in \eqref{eqn:sigmaFinal} as 
\begin{equation}\label{eqn:sigmaFinalDenom}
\frac{\sigma^{2}}{\sigma_{0}^{2}} + \sum_{t=0}^{T-1} h(\lambda_{i}(t)) = \frac{\sigma^{2}}{\sigma_{i}^{2}(T-1)} + h(\lambda_{i}(T-1)).
\end{equation}
%
We then decompose 
the expectation in \eqref{eqn:costFun} into an expectation over $\mby(T)$ conditioned on $\mbY(T-1)$ followed by an expectation over $\mbY(T-1)$.  Note that only $p_{i}(T)$ depends on $\mby(T)$ in \eqref{eqn:costFun}. 
Taking the expectation of \eqref{eqn:IpostRec} with respect to $\mby(t) \mid \mbY(t-1)$ yields 
\begin{equation}\label{eqn:pExp}
\E \{p_{i}(t) \mid \mbY(t-1)\} = p_{i}(t-1), \quad t = 1, \ldots, T.
\end{equation}
%
Using \eqref{eqn:sigmaFinal}, \eqref{eqn:sigmaFinalDenom}, and \eqref{eqn:pExp}, the effort allocation problem may be stated as 
\begin{equation}\label{eqn:prob}
\begin{split}
\min_{\mblambda(0), \ldots, \mblambda(T-1)} \quad &\E \left\{ G(\mbx(T-1), \mblambda(T-1)) \right\} \\
\text{s.t.} \quad &\sum_{t=0}^{T-1} \sum_{i=1}^{N} \lambda_{i}(t) = \Lambda_{0}, \quad \lambda_{i}(t) \geq 0 \;\; \forall \; t, i,
\end{split}
\end{equation}
%
%
%
where the cost function is of the desired form with a single non-zero term at time $T-1$,
\ifCLASSOPTIONdraftcls
\begin{align}
G(\mbx(T-1), \mblambda(T-1)) &= 
\sum_{i=1}^{N} p_{i}(T-1) g(\sigma_{i}^{2}(T-1), h(\lambda_{i}(T-1))),\label{eqn:costT-1}\\
g(\sigma_{i}^{2}(t), \hb_{i}) &= \int_{0}^{\infty} L\left( \frac{\sigma \theta}{\sqrt{\sigma^{2} / \sigma_{i}^{2}(t) + \hb_{i} }} \right) \phi(\theta) \, d\theta, \label{eqn:gi}
\end{align}
\else
\begin{gather}
\begin{split}
&G(\mbx(T-1), \mblambda(T-1)) = \\
&\qquad\qquad\sum_{i=1}^{N} p_{i}(T-1) g(\sigma_{i}^{2}(T-1), h(\lambda_{i}(T-1))),\label{eqn:costT-1}
\end{split}\\
g(\sigma_{i}^{2}(t), \hb_{i}) = \int_{0}^{\infty} L\left( \frac{\sigma \theta}{\sqrt{\sigma^{2} / \sigma_{i}^{2}(t) + \hb_{i} }} \right) \phi(\theta; 0,1) \, d\theta, \label{eqn:gi}
\end{gather}
\fi
depending explicitly on $\mbp(T-1)$, $\mbsigma^{2}(T-1)$, and $\mblambda(T-1)$.  The dependence on the variables $\mblambda(t)$, $t = 0, \ldots, T-2$ is implicit through the probability distribution of the observations $\mbY(T-1)$ and the recursions in \eqref{eqn:stateEvol}.  The constraints in \eqref{eqn:prob} actually represent a continuum of constraints since they are required to be satisfied for all realizations of $\mbY(T-1)$.

\section{Effort allocation policies}
\label{sec:policy}

In this section, we develop policies directed at solving the effort allocation problem \eqref{eqn:prob}.  Optimal policies are discussed in Section \ref{subsec:policyOpt} while a less complex method known as open-loop feedback control is discussed in Section \ref{subsec:policyOLFC}.  We then discuss two approaches to improving the performance of OLFC: generalized OLFC in Section \ref{subsec:policyOLFCGen}, and policy rollout in Section \ref{subsec:policyOLFCRollout}.

\subsection{Optimal policies}
\label{subsec:policyOpt}

In principle, it is possible to employ exact dynamic programming to determine an optimal policy for \eqref{eqn:prob}.  The dynamic programming approach decomposes \eqref{eqn:prob} into a sequence of optimizations proceeding backward in time, making 
repeated use of iterated expectations and the fact that each allocation $\mblambda(t)$ is a function of past observations $\mbY(t)$ but not future ones.  The last-stage optimization is given by 
\begin{subequations}
\begin{equation}\label{eqn:JT-1*}
\begin{split}
J_{T-1}^{\ast}(\mbx(T-1)) = \min_{\mblambda(T-1)} \quad &G(\mbx(T-1),\mblambda(T-1)) \\
\text{s.t.} \quad &\sum_{i=1}^{N} \lambda_{i}(T-1) = \Lambda(T-1), \\ 
&\lambda_{i}(T-1) \geq 0 \quad \forall \; i,
\end{split}
\end{equation}
and for $t = T-2, T-3, \ldots, 0$, the optimizations are defined recursively as follows: 
\begin{equation}\label{eqn:Jt*}
\begin{split}
J_{t}^{\ast}(\mbx(t)) = \min_{\mblambda(t)} \quad &\E \left\{ J_{t+1}^{\ast}(\mbx(t+1)) \mid \mbx(t), \mblambda(t) \right\} \\
\text{s.t.} \quad &\sum_{i=1}^{N} \lambda_{i}(t) \leq \Lambda(t), \quad \lambda_{i}(t) \geq 0 \;\; \forall \; i.
\end{split}
\end{equation}
\end{subequations}
The functions $J_{t}^{\ast}(\mbx(t))$ represent the optimal costs-to-go starting from stage $t$ and state $\mbx(t)$, and thus the desired optimal cost in \eqref{eqn:prob} is $J_{0}^{\ast}(\mbx(0))$.  The notation in \eqref{eqn:Jt*} reflects the fact that the distribution of $\mby(t+1)$ given $\mbY(t)$ is completely determined by $\mbx(t)$ and $\mblambda(t)$; more specifically, $f(y_{i}(t+1) \mid \mbY(t))$ is given by the denominator of the right-hand side of \eqref{eqn:pEvol} as can be seen from \eqref{eqn:IiPostRec}.  The next state $\mbx(t+1)$ is specified by $\mbx(t)$, $\mblambda(t)$, and $\mby(t+1)$ through \eqref{eqn:stateEvol}.  Thus the choice of $\mblambda(t)$ depends on $\mbY(t)$ only through the state $\mbx(t)$, which is a property of dynamic programs \cite{bertsekas2005}.

An optimal policy can be obtained by first solving \eqref{eqn:JT-1*} for $\mblambda(T-1)$ and then using the result in \eqref{eqn:Jt*} to solve for $\mblambda(T-2)$.  The remaining allocations are determined in the same recursive way.  This exact procedure is computationally tractable only in a few cases.  For $T = 1$, it suffices to solve \eqref{eqn:JT-1*}, which is a convex optimization problem under some conditions to be discussed in Section \ref{subsec:policyOLFC}.  For $T = 2$ and a uniform prior ($p_{i}(0) = p_{0}$, $\mu_{i}(0) = \mu_{0}$, $\sigma_{i}^{2}(0) = \sigma_{0}^{2}$), symmetry allows the initial allocation $\mblambda(0)$ to be restricted to the form $\mblambda(0) = \beta^{(2)}(0) \ones$, where $\ones$ denotes a vector with unit entries.  Thus \eqref{eqn:Jt*} becomes a one-dimensional optimization with respect to the multiplier $\beta^{(2)}(0)$.  For fixed $\beta^{(2)}(0)$, the expectation in \eqref{eqn:Jt*} can be evaluated by sampling from the distribution of $\mby(1)$ and then solving \eqref{eqn:JT-1*} for the resulting values of the state $\mbx(1)$.  

For $T > 2$ however, an exact solution via \eqref{eqn:JT-1*} and \eqref{eqn:Jt*} is very difficult.  
The first issue is that the objective function in \eqref{eqn:Jt*} is defined recursively in terms of $J_{t+1}^{\ast}(\mbx(t+1))$ and the high dimension and continuous nature of the state make it difficult to summarize $J_{t+1}^{\ast}(\mbx(t+1))$ by storing its values at a small number of representative states $\mbx(t+1)$.  Second, even if the objective function could be readily computed, each evaluation of \eqref{eqn:Jt*} involves in general an $N$-dimensional optimization with no known structure and $N$ potentially large.
For these reasons, we do not consider an exact solution to \eqref{eqn:prob} for $T > 2$, opting instead for an approximate method as is discussed next.

\subsection{Open-loop feedback control}
\label{subsec:policyOLFC}

A well-known approach to approximate dynamic programming is that of open-loop feedback control (OLFC) \cite{bertsekas2005}.  
We consider the problem of determining the allocation $\mblambda(t)$ at time $t$ given the current set of observations $\mbY(t)$, or equivalently the state $\mbx(t)$.  In OLFC, this computation is simplified by assuming that future allocations $\mblambda(t+1),\ldots,\mblambda(T-1)$ can depend only on $\mbY(t)$ and not future observations.  In other words, planning for future allocations is done open-loop.  Once the allocations $\mblambda(t), \ldots, \mblambda(T-1)$ are determined, the first allocation $\mblambda(t)$ is used to obtain new observations $\mby(t+1)$ and the state is updated to $\mbx(t+1)$.  The allocations $\mblambda(t+1), \ldots, \mblambda(T-1)$ are then recomputed, this time based on $\mbx(t+1)$ and under the same assumption regarding the future $t+2,\ldots,T$.

In light of the OLFC assumption, the only quantities that depend on $\mby(t+1),\ldots,\mby(T-1)$ in \eqref{eqn:costT-1} are the probabilities $p_{i}(T-1)$.  The conditional expectations 
with respect to $\mby(T-1) \mid \mbY(T-2), \mby(T-2) \mid \mbY(T-3), \ldots, \mby(t+1) \mid \mbY(t)$ in \eqref{eqn:prob} 
can then be applied to transform $p_{i}(T-1)$ into $p_{i}(t)$ using \eqref{eqn:pExp} repeatedly.  The resulting cost function is to be optimized with respect to $\mblambda(t), \ldots, \mblambda(T-1)$ jointly, leading to the problem 
%
\begin{equation}\label{eqn:probOLFC}
\begin{split}
\min_{\mblambda(t),\ldots,\mblambda(T-1)} \quad &\sum_{i=1}^{N} p_{i}(t) g\left(\sigma_{i}^{2}(t), \sum_{\tau=t}^{T-1} h(\lambda_{i}(\tau)) \right)\\
\text{s.t.} \quad &\sum_{\tau=t}^{T-1} \sum_{i=1}^{N} \lambda_{i}(\tau) = \Lambda(t), 
\quad \lambda_{i}(\tau) \geq 0 \;\; \forall \; \tau, i,
\end{split}
\end{equation}
where we have made use of a rearrangement similar to \eqref{eqn:sigmaFinalDenom}.  The budget constraint in \eqref{eqn:probOLFC} is assumed to be met with equality as otherwise the cost could be decreased.  

For $t = T-1$, the OLFC problem \eqref{eqn:probOLFC} coincides with the last-stage optimization in \eqref{eqn:JT-1*}.  For $t < T-1$, OLFC represents a significant simplification relative to the exact optimization in \eqref{eqn:Jt*} because the cost function in \eqref{eqn:probOLFC} is expressed explicitly without the need to evaluate expectations recursively.  Under certain conditions specified in the following proposition, problem \eqref{eqn:probOLFC} is also a convex optimization and thus can be tractably solved.

\begin{prop}\label{prop:OLFCconvex}
The OLFC problem \eqref{eqn:probOLFC} is a convex optimization problem if the loss function $L$ is non-decreasing, $g(\sigma_{i}^{2}(t), \hb_{i})$ in \eqref{eqn:gi} 
%
%
is a convex function of $\hb_{i}$ for $\hb_{i} \geq 0$ and all $\sigma_{i}^{2}(t)$, and the effort function $h$ is concave.
\end{prop}
\begin{IEEEproof}
Since the constraints in \eqref{eqn:probOLFC} are all linear, the feasible set is convex (more precisely a simplex).  The cost function is a non-negative combination of functions $g(\sigma_{i}^{2}(t), \hb_{i})$ with $\hb_{i} = \sum_{\tau=t}^{T-1} h(\lambda_{i}(\tau))$, so it suffices to prove that $g$ is convex as a function of $\lambda_{i}(t),\ldots,\lambda_{i}(T-1)$.  First note that $\hb_{i}$, as a sum of concave functions, is concave in $\lambda_{i}(t),\ldots,\lambda_{i}(T-1)$.  Given that $L$ is a non-decreasing function of its argument, $g$ is seen to be a non-increasing function of $\hb_{i}$.  Furthermore, we may extend the definition of $g$ to negative $\hb_{i}$ by letting $g(\sigma_{i}^{2}(t), \hb_{i}) = \infty$ for $\hb_{i} < 0$, thereby preserving the monotonicity and assumed convexity of $g$.  It then follows from a property of compositions of functions \cite{bv2004} that $g$ is convex in $\lambda_{i}(t),\ldots,\lambda_{i}(T-1)$.
\end{IEEEproof}

The assumptions in Proposition \ref{prop:OLFCconvex} are not difficult to satisfy.  It was already assumed in Section \ref{sec:prob} that $L$ is non-decreasing so that the optimal amplitude estimate $\thetah_{i}$ is equal to the conditional mean $\mu_{i}(T)$.  The concavity assumption on $h$ is satisfied by the identity function as well as functions corresponding to a sublinear dependence of the observation precision on sensing effort.  The convexity assumption on $g$ is satisfied by a variety of commonly used loss functions.  As a first example we consider the $0$-$1$ loss function for a tolerance $\epsilon$,
\[
L_{\epsilon}(a) = \begin{cases}
0, & 0 \leq a < \epsilon,\\
1, & a > \epsilon.
\end{cases}
\]
The integral in \eqref{eqn:gi} may be evaluated in this case to yield 
\[
g(\sigma_{i}^{2}(t), \hb_{i}) = Q\left( \frac{\epsilon}{\sigma} \sqrt{\sigma^{2}/\sigma_{i}^{2}(t) + \hb_{i}} \right),
\]
where $Q$ denotes the Q-function, i.e., the standard Gaussian tail probability.  Since the Q-function is convex decreasing for non-negative arguments and the square root function is concave in $\hb_{i}$, the same property used in the proof of Proposition \ref{prop:OLFCconvex} may be invoked to conclude that $g$ is a convex function of $\hb_{i}$.  
%
%
The convexity of $g$ can also be verified for $L(a) = 1 - e^{-ba}$ with $b > 0$, which can be regarded as a continuous approximation to the $0$-$1$ loss function.

The assumption that $g$ is convex may be replaced by one of the following stricter but more easily checked conditions:
\begin{enumerate}
\item[(a)] $L(1 / \sqrt{h})$ is a convex function of $h$;
\item[(b)] $L$ is convex.
\end{enumerate}
Condition (a) implies that $g$ is convex because shifting and scaling the argument of a function do not affect convexity and because the weighting function $\phi(\theta; 0, 1)$ in the integral in \eqref{eqn:gi} is always positive.  Condition (b) implies condition (a) because of a composition property similar to the one used earlier and the convexity of $1/\sqrt{h}$ with respect to $h$.  If $L$ is twice differentiable, condition (a) can be shown to be equivalent to the inequality  
\begin{equation}\label{eqn:Lconvex}
a L''(a) + 3 L'(a) \geq 0, \qquad a \geq 0,
\end{equation}
whereas (b) is equivalent to $L''(a) \geq 0$.  Condition (b) includes the square loss $L(a) = a^{2}$ corresponding to MSE, the linear loss $L(a) = a$ corresponding to mean absolute error (MAE), the Huber loss which combines the square and linear losses in a continuous and convex manner, and the two-sided hinge loss.  More generally, \eqref{eqn:Lconvex} is satisfied for any power-law function $L(a) = a^{q}$ with $q > 0$ and for $L(a) = \log(1 + ba)$ with $b > 0$.  Note that $a^{q}$ for $0 < q < 1$ and $\log(1 + ba)$ are concave functions of $a$.  Taking the limit as $q \to 0$ of the power-law functions yields the $0$-$1$ loss function, which was shown earlier to result in a convex $g$.

In the remainder of the paper, we assume that the assumptions of Proposition \ref{prop:OLFCconvex} are satisfied and hence the OLFC problem \eqref{eqn:probOLFC} is a convex optimization.  We now address the solution of \eqref{eqn:probOLFC}.  The cost function in \eqref{eqn:probOLFC} depends on $\mblambda(t),\ldots,\mblambda(T-1)$ only through the quantities $\hb_{i} = \sum_{\tau=t}^{T-1} h(\lambda_{i}(\tau))$, and is more specifically a non-increasing function of $\hb_{i}$ as argued in the proof of Proposition \ref{prop:OLFCconvex}.  Therefore \eqref{eqn:probOLFC} may be solved via a two-step procedure: first we fix $\lambdab_{i}(t) = \sum_{\tau=t}^{T-1} \lambda_{i}(\tau)$ and seek to maximize $\hb_{i}$ as functions of $\lambdab_{i}(t)$, i.e., 
\begin{equation}\label{eqn:hbar*}
\begin{split}
\hb_{i}^{\ast}(\lambdab_{i}(t)) = \max_{\lambda_{i}(t),\ldots,\lambda_{i}(T-1)} \quad &\sum_{\tau=t}^{T-1} h(\lambda_{i}(\tau)) \\ 
\text{s.t.} \quad &\sum_{\tau=t}^{T-1} \lambda_{i}(\tau) = \lambdab_{i}(t),\\
&\lambda_{i}(\tau) \geq 0 \quad \forall \; \tau, i,
\end{split}
\end{equation}
and then we substitute the maximum values $\hb_{i}^{\ast}(\lambdab_{i}(t))$ into \eqref{eqn:probOLFC} and optimize with respect to $\lambdab_{i}(t)$.  The maximum $\hb_{i}^{\ast}(\lambdab_{i}(t))$ can be determined by noting that \eqref{eqn:hbar*} is a concave maximization problem subject to a simplex constraint.  For such problems, we have the following necessary and sufficient optimality condition: 
\begin{equation}\label{eqn:optCondSimplexMax}
\text{if} \quad \lambda_{i}^{\ast}(\tau) > 0 \quad 
\text{then} \quad \frac{\partial \hb_{i}}{\partial\lambda_{i}(\tau)} \geq \frac{\partial \hb_{i}}{\partial\lambda_{i}(\tau')} \quad \forall \; \tau' \neq \tau,
\end{equation}
where the partial derivatives are evaluated at the optimum.  The solution $\lambda_{i}^{\ast}(\tau) = \lambdab_{i}(t) / (T-t)$ for all $\tau$ satisfies \eqref{eqn:optCondSimplexMax} by symmetry since all of the partial derivatives are equal.  The corresponding maximum value is therefore $\hb_{i}^{\ast}(\lambdab_{i}(t)) = (T-t) h(\lambdab_{i}(t) / (T-t))$.  Note however that the optimal solution to \eqref{eqn:hbar*} may not be unique if $h$ is not strictly concave.  In particular, if $h$ is the identity function, then $\hb_{i} = \lambdab_{i}(t)$ regardless of the choice of $\lambda_{i}(t),\ldots,\lambda_{i}(T-1)$.  We return to the issue of non-uniqueness in Section \ref{subsec:policyOLFCGen}.

With the substitutions $\sum_{\tau=t}^{T-1} \lambda_{i}(\tau) = \lambdab_{i}(t)$ and $\sum_{\tau=t}^{T-1} h(\lambda_{i}(\tau)) = (T-t) h(\lambdab_{i}(t) / (T-t))$, \eqref{eqn:probOLFC} simplifies to 
\begin{equation}\label{eqn:probOLFCBar}
\begin{split}
\min_{\mblambdab(t)} \quad &\sum_{i=1}^{N} p_{i}(t) g\left( \sigma_{i}^{2}(t), (T-t) h\left( \frac{\lambdab_{i}(t)}{T-t} \right) \right) \\
\text{s.t.} \quad &\sum_{i=1}^{N} \lambdab_{i}(t) = \Lambda(t), 
\quad \lambdab_{i}(t) \geq 0 \;\; \forall \; i,
\end{split}
\end{equation}
a simplex-constrained convex minimization problem.  Problem \eqref{eqn:probOLFCBar} thus satisfies an optimality condition similar to \eqref{eqn:optCondSimplexMax} with the inequality between partial derivatives reversed in direction.  This condition implies that optimal solutions to \eqref{eqn:probOLFCBar} have certain properties akin to water-filling.  First, the solutions exhibit thresholding in the sense that $\lambdab_{i}^{\ast}(t)$ must be zero if the corresponding partial derivative is not among the lowest.  Second, the partial derivatives corresponding to non-zero components must all be equal.  This in turn induces an ordering among the non-zero allocations as a function of the probabilities $p_{i}(t)$ and variances $\sigma_{i}^{2}(t)$.
%
%

To illustrate the properties of optimal solutions to \eqref{eqn:probOLFCBar}, we specialize to the case of power-law losses $L(a) = a^{q}$ and the identity effort function $h(\lambda) = \lambda$.  In this case, \eqref{eqn:probOLFCBar} reduces to 
\begin{equation}\label{eqn:probOLFCMSE}
\begin{split}
\min_{\mblambdab(t)} \quad &\sum_{i=1}^{N} \frac{p_{i}(t)}{(\sigma^{2} / \sigma_{i}^{2}(t) + \lambdab_{i}(t))^{q/2}} \\
\text{s.t.} \quad &\sum_{i=1}^{N} \lambdab_{i}(t) = \Lambda(t), 
\quad \lambdab_{i}(t) \geq 0 \;\; \forall \; i,
\end{split}
\end{equation}
and the optimal solution can be stated explicitly.  A detailed derivation is provided in Appendix \ref{app:OLC}.  First we define $\gamma = 2 / (q + 2)$ and $\pi$ to be an index permutation that sorts the quantities $p_{i}^{\gamma}(t) \sigma_{i}^{2}(t)$ in non-increasing order:
\begin{equation}\label{eqn:orderingOLFC}
p_{\pi(1)}^{\gamma}(t) \sigma_{\pi(1)}^{2}(t) \geq p_{\pi(2)}^{\gamma}(t) \sigma_{\pi(2)}^{2}(t) \geq \dots \geq p_{\pi(N)}^{\gamma}(t) \sigma_{\pi(N)}^{2}(t).
\end{equation}
Next define $b(k)$ to be the monotonically non-decreasing function of $k = 0, 1, \ldots, N$ with $b(N) = \infty$ and 
\ifCLASSOPTIONdraftcls
\begin{equation}\label{eqn:bOLFC}
b(k) = \frac{\sigma^{2}}{p_{\pi(k+1)}^{\gamma}(t) \sigma_{\pi(k+1)}^{2}(t)} \sum_{i=1}^{k} p_{\pi(i)}^{\gamma}(t) - \sum_{i=1}^{k} \frac{\sigma^{2}}{\sigma_{\pi(i)}^{2}(t)}, \qquad k = 0, \ldots, N-1.
\end{equation}
\else
\begin{multline}\label{eqn:bOLFC}
b(k) = \frac{\sigma^{2}}{p_{\pi(k+1)}^{\gamma}(t) \sigma_{\pi(k+1)}^{2}(t)} \sum_{i=1}^{k} p_{\pi(i)}^{\gamma}(t) - \sum_{i=1}^{k} \frac{\sigma^{2}}{\sigma_{\pi(i)}^{2}(t)},\\ k = 0, \ldots, N-1.
\end{multline}
\fi
Then the optimal solution $\mblambdab^{\ast}(t)$ to \eqref{eqn:probOLFCMSE} is given by 
\begin{equation}\label{eqn:lambdab*}
\lambdab_{\pi(i)}^{\ast}(t) = \begin{cases}
C p_{\pi(i)}^{\gamma}(t) - \frac{\sigma^{2}}{\sigma_{\pi(i)}^{2}(t)}, & i = 1, \ldots, k,\\
0, & i = k+1, \ldots, N,
\end{cases}
\end{equation}
where 
\begin{equation}\label{eqn:C}
C = \frac{\Lambda(t) + \sum_{j=1}^{k} \frac{\sigma^{2}}{\sigma_{\pi(j)}^{2}(t)}}{\sum_{j=1}^{k} p_{\pi(j)}^{\gamma}(t)}
\end{equation}
and the number of non-zero components $k$ is determined by the interval $(b(k-1), b(k)]$ to which the budget parameter $\Lambda(t)$ belongs.
The monotonicity of $b(k)$ ensures that the mapping from $\Lambda(t)$ to $k$ is well-defined.  We note that $k$ and $C$ could also be computed using the general procedure in \cite{palomar2005}.  The thresholding property is clearly seen in \eqref{eqn:lambdab*}.  Furthermore, the non-zero allocations increase with the probabilities $p_{i}(t)$ raised to the power $\gamma$ and decrease with the precisions $1/\sigma_{i}^{2}(t)$.  

In the case of general loss and effort functions, \eqref{eqn:probOLFCBar} may not have an explicit solution as in \eqref{eqn:orderingOLFC}--\eqref{eqn:C}.  Nevertheless, an efficient iterative solution is possible under the assumption of convexity.  One possibility is to use a projected gradient algorithm, taking advantage of the ease of projecting onto a simplex.

The solution to \eqref{eqn:probOLFCBar} specifies the values of the sums $\lambdab_{i}(t) = \sum_{\tau=t}^{T-1} \lambda_{i}(\tau)$.  However, the solution to \eqref{eqn:hbar*} may not uniquely specify the division of $\lambdab_{i}(t)$ into $\lambda_{i}(t), \ldots, \lambda_{i}(T-1)$ if the effort function $h$ is not strictly concave.  
In addition, since the OLFC optimization \eqref{eqn:probOLFCBar} is similar to the last-stage optimization \eqref{eqn:JT-1*}, the resulting policy can be somewhat aggressive in allocating effort to components currently believed to contain signal as opposed to waiting for further confirmation.  In the next subsection, these issues are addressed through a generalization of the OLFC approach.

\subsection{Generalized open-loop feedback control}
\label{subsec:policyOLFCGen}

In this subsection, we discuss two modifications to the OLFC policy in Section \ref{subsec:policyOLFC}.  The first modification is directed at optimizing the distribution of effort over stages and applies to all loss functions.  The second modification reduces premature exploitation and is presented only for power-law loss functions;  similar strategies could be devised for other loss functions.  
As seen in Proposition \ref{prop:nested} below, the modifications ensure that the resulting policies improve monotonically with the number of stages $T$.

To optimize the allocation of effort over stages, we restrict the allocation for the current stage $\mblambda(t)$ to be proportional to the optimal solution $\mblambdab^{\ast}(t)$ of \eqref{eqn:probOLFCBar}, i.e., $\mblambda(t) = \beta^{(T)}(t) \mblambdab^{\ast}(t)$, where $\beta^{(T)}(t) \in [0, 1]$ represents the fraction of the remaining budget $\Lambda(t)$ that is used at time $t$ and the superscript $T$ denotes the total number of stages.  The fractions $\beta^{(T)}(t)$ are chosen based on a generalization of the optimal policies for $T = 1$ and $T = 2$ in Section \ref{subsec:policyOpt}.  Both of these optimal policies belong to the OLFC class.  Specifically, the $T = 1$ policy results from solving \eqref{eqn:JT-1*}, which is a special case of \eqref{eqn:probOLFCBar} with $t = T-1$, and setting $\beta^{(1)}(0) = 1$ since there is only one stage.  The $T = 2$ policy uses an initial allocation $\mblambda(0) = \beta^{(2)}(0) \ones$, which is of the same form as the solution to \eqref{eqn:probOLFCBar} for $t = 0$ under a uniform prior, followed by the solution to \eqref{eqn:probOLFCBar} for $t = 1$ scaled by $\beta^{(2)}(1) = 1$.  
Note that the second stage in the $T = 2$ policy is identical to the $T = 1$ policy with $\beta^{(2)}(1) = \beta^{(1)}(0)$.  For $T > 2$, we follow the same strategy of reusing the $(T-1)$-stage fractions in the $T$-stage policy, setting $\beta^{(T)}(t) = \beta^{(T-1)}(t-1)$ for $t = 1, 2, \ldots, T-1$. 
The first-stage fraction $\beta^{(T)}(0)$ is then optimized as described below.  

The second modification is to allow the exponent $\gamma$ in \eqref{eqn:orderingOLFC}--\eqref{eqn:C} to vary with time.  The last-stage exponent $\gamma^{(T)}(T-1)$ is set to $2/(q+2)$, the optimal exponent for the loss function $L(a) = a^{q}$.  In earlier stages, smaller exponents are used to make the policy more conservative, specifically by weakening the dependence on the probabilities $p_{i}(t)$.  We propose the simple strategy of optimizing only the first-stage exponent $\gamma^{(T)}(0)$ and constraining the remaining exponents to linearly interpolate between $\gamma^{(T)}(0)$ and $\gamma^{(T)}(T-1)$.  This reduces the determination of the fractions $\beta^{(T)}(t)$ and exponents $\gamma^{(T)}(t)$ to a two-dimensional optimization regardless of the number of stages.

The first-stage parameters $\beta^{(T)}(0)$ and $\gamma^{(T)}(0)$ are determined recursively for $T = 1, 2, \ldots$ starting from $\beta^{(1)}(0) = 1$ and $\gamma^{(1)}(0) = 2/(q+2)$.  Define $J_{t}^{(T)}(\mbx(t))$ to be the cost-to-go of a $T$-stage policy in this family starting from time $t$ and state $\mbx(t)$.  Then for $T > 1$, $\beta^{(T)}(0)$ and $\gamma^{(T)}(0)$ are given by  
\ifCLASSOPTIONdraftcls
\begin{equation}\label{eqn:beta0Nested}
\left( \beta^{(T)}(0), \gamma^{(T)}(0) \right) 
= \argmin_{\substack{0 \leq \beta \leq 1\\\gamma \leq 2/(q+2)}} \quad \textcolor{red}{\E} \left\{ J_{1}^{(T)}(\mbx(1)) \mid \mbx(0), \beta\mblambdab^{\ast}(0) \right\}.
\end{equation}
\else
\begin{multline}\label{eqn:beta0Nested}
\left( \beta^{(T)}(0), \gamma^{(T)}(0) \right)\\ 
= \argmin_{\substack{0 \leq \beta \leq 1\\\gamma \leq 2/(q+2)}} \quad \E \left\{ J_{1}^{(T)}(\mbx(1)) \mid \mbx(0), \beta\mblambdab^{\ast}(0) \right\}.
\end{multline}
\fi
%
The parameters $\beta^{(T)}(1), \ldots, \beta^{(T)}(T-2)$ and $\gamma^{(T)}(1), \ldots, \gamma^{(T)}(T-2)$ required to evaluate $J_{1}^{(T)}$ are specified by the $(T-1)$-stage policy and the choice of $\gamma$.  The expectation in \eqref{eqn:beta0Nested} can be computed by sampling from the distribution of $\mby(1)$, determining the state $\mbx(1)$ using \eqref{eqn:stateEvol}, and then simulating the remainder of the policy.  All of these computations can be done offline since they depend only on the initial state $\mbx(0)$ and previously determined policies.  In addition, since the optimization in \eqref{eqn:beta0Nested} can partially account for the effect of future observations on future allocations, an effect that is ignored in the OLFC simplification, the optimization over $\beta$ is performed even in the case of strictly concave $h$.  Otherwise, \eqref{eqn:hbar*} would yield a uniform distribution over stages corresponding to $\beta^{(T)}(t) = 1/(T-t)$.  

The family of generalized OLFC policies defined above satisfies the following monotonic improvement property. 
%
\begin{prop}\label{prop:nested}
The cost of the generalized OLFC policies is non-increasing in the number of stages, i.e., 
\[
J_{0}^{(T)}(\mbx(0)) \leq J_{0}^{(T-1)}(\mbx(0)), \quad T = 2, 3, \ldots.
\]
\end{prop}
\begin{IEEEproof}
The cost of the $T$-stage policy is given by 
\begin{equation}\label{eqn:costNested}
J_{0}^{(T)}(\mbx(0)) = \min_{\substack{0 \leq \beta \leq 1\\\gamma \leq 2/(q+2)}}  \E \left\{ J_{1}^{(T)}(\mbx(1)) \mid \mbx(0), \beta\mblambdab^{\ast}(0) \right\}.
\end{equation}
Consider fixing $\beta = 0$ and 
\[
\gamma = \begin{cases}
\frac{2}{q+2}, & T = 2,\\
\frac{T-1}{T-2} \gamma^{(T-1)}(0) - \frac{1}{T-2} \frac{2}{q+2}, & T > 2
\end{cases}
\]
on the right-hand side of \eqref{eqn:costNested}.  With $\beta = 0$, the observations $\mby(1)$ are not taken, the state $\mbx(1)$ is unchanged from $\mbx(0)$, and the budget usage fractions are the same as in the $(T-1)$-stage policy.  It can also be seen from the choice of $\gamma$ that the exponents are the same as for $T-1$, and hence the right-hand side of \eqref{eqn:costNested} reduces to $J_{0}^{(T-1)}(\mbx(0))$.  The claim then follows.
\end{IEEEproof}
Proposition \ref{prop:nested} implies in particular that the generalized OLFC policies for $T > 2$ improve upon the optimal policy for $T = 2$.  The corresponding performance gains 
are quantified through numerical simulations in Section \ref{sec:perfEval}.


\subsection{Rollout OLFC policies}
\label{subsec:policyOLFCRollout}


We now discuss a different approach to improving the performance of OLFC based on the dynamic programming technique of \emph{policy rollout} \cite{bertsekas2005}. 
For simplicity, we assume that the exponent $\gamma$ in \eqref{eqn:orderingOLFC}--\eqref{eqn:C} is fixed to $2/(q+2)$ in all stages, unlike in Section \ref{subsec:policyOLFCGen}.  In this subsection only, we also make the same assumption for the generalized OLFC policies, i.e., the only parameter optimized in \eqref{eqn:beta0Nested} is $\beta^{(T)}(0)$.  Rollout could also be applied in the case of time-varying $\gamma$ by changing the optimization over $\beta(t)$ in \eqref{eqn:betaRollout} below to a joint optimization over $\beta(t)$ and $\gamma(t)$.

In the last stage $t = T-1$ of a rollout policy, the allocation is determined as before by solving \eqref{eqn:JT-1*}, or equivalently by solving \eqref{eqn:probOLFC} with budget usage fraction $\betat^{(T)}(T-1) = 1$ (we use a tilde to distinguish the rollout fractions from those in the generalized OLFC policies).  For $t = 0, 1, \ldots, T-2$, the fraction $\betat^{(T)}(t)$ is determined according to 
\begin{equation}\label{eqn:betaRollout}
\betat^{(T)}(t) = \arg\min_{0 \leq \beta(t) \leq 1} \quad \E \left\{ J_{t+1}^{(T)}(\mbx(t+1)) \mid \mbx(t), \beta(t)\mblambdab^{\ast}(t) \right\},
\end{equation}
%
where $J_{t+1}^{(T)}(\mbx(t+1))$ is the cost-to-go of the $T$-stage generalized policy.  Thus $\betat^{(T)}(t)$ is chosen assuming that future stages follow the generalized policy.  The corresponding cost-to-go $J_{t+1}^{(T)}(\mbx(t+1))$ can be viewed as an approximation to the optimal cost-to-go $J_{t+1}^{\ast}(\mbx(t+1))$ in \eqref{eqn:Jt*}.  Comparing \eqref{eqn:betaRollout} with \eqref{eqn:beta0Nested} 
(and assuming that $\gamma(t) = 2/(q+2)$ for all $t$), it is seen that $\betat^{(T)}(0) = \beta^{(T)}(0)$.  In other stages however, the rollout fractions differ from those in the corresponding generalized policy because they are re-optimized based on the value of the current state $\mbx(t)$ instead of being taken directly from a policy with fewer stages.  

In general, rollout policies have the property of improved performance over the policies on which they are based.  The same holds for the present rollout policy, with the difference being that the optimization in \eqref{eqn:betaRollout} is restricted to a line search over $\beta(t)$.  Denoting by $\Jt_{t}^{(T)}(\mbx(t))$ the cost-to-go of a $T$-stage rollout policy starting from time $t$ and state $\mbx(t)$, we have the following result:
\begin{prop}\label{prop:rollout}
The $T$-stage rollout OLFC policy has a lower cost-to-go than the corresponding generalized OLFC policy in all stages and states, i.e., 
\[
\Jt_{t}^{(T)}(\mbx(t)) \leq J_{t}^{(T)}(\mbx(t)), \quad t = 0, \ldots, T-1, \quad \forall \; \mbx(t).
\]
\end{prop}  
\begin{IEEEproof}
The proof is based on \cite[Sec. 6.4]{bertsekas2005}.  For $t = T-1$, the two policies coincide so the costs-to-go are the same.  Assume inductively that $\Jt_{t+1}^{(T)}(\mbx(t+1)) \leq J_{t+1}^{(T)}(\mbx(t+1))$ for all $\mbx(t+1)$.  The cost-to-go of the rollout policy is defined by 
\[
\Jt_{t}^{(T)}(\mbx(t)) = \E \left\{ \Jt_{t+1}^{(T)}(\mbx(t+1)) \mid \mbx(t), \betat^{(T)}(t)\mblambdab^{\ast}(t) \right\}
\]
and similarly for the nested policy.  By the induction hypothesis and the definition of the rollout policy \eqref{eqn:betaRollout},
\begin{align*}
\Jt_{t}^{(T)}(\mbx(t)) &\leq \E \left\{ J_{t+1}^{(T)}(\mbx(t+1)) \mid \mbx(t), \betat^{(T)}(t)\mblambdab^{\ast}(t) \right\}\\
&\leq \E \left\{ J_{t+1}^{(T)}(\mbx(t+1)) \mid \mbx(t), \beta^{(T)}(t)\mblambdab^{\ast}(t) \right\}\\
&= J_{t}^{(T)}(\mbx(t))
\end{align*}
for all $\mbx(t)$ as required.  Note that the second inequality depends on the generalized policy being included in the class over which the rollout policy is optimized.
\end{IEEEproof}

The rollout OLFC policies can make greater use of knowledge of the state $\mbx(t)$ but are consequently more demanding computationally than the generalized OLFC policies.  Instead of a single optimization in \eqref{eqn:beta0Nested}, $T-1$ optimizations as in \eqref{eqn:betaRollout} are required.  Furthermore and in contrast to \eqref{eqn:beta0Nested}, \eqref{eqn:betaRollout} must be evaluated online since it depends on the current state $\mbx(t)$.  The simulations involved in computing the expectation in \eqref{eqn:betaRollout} do become shorter however as $t$ increases.  The improvement due to rollout is characterized through numerical simulations in Section \ref{sec:perfEval}.

\section{Numerical simulations}
\label{sec:perfEval}

Numerical simulations are used to evaluate the OLFC policies developed in Section \ref{sec:policy}.  The monotonic improvement property of Proposition \ref{prop:nested} is verified and gains up to several dB are observed relative to the optimal two-stage policy.  The proposed policies are also seen to consistently outperform distilled sensing (DS), most significantly at higher SNR.  
We have additionally made comparisons to the sequential thresholding method in \cite{malloy2011a,malloy2011b}, which in the case of Gaussian observations is similar to DS except in its allocation of sensing effort over stages.  In terms of estimation loss \eqref{eqn:expLossROI}, we find that DS performs uniformly better than sequential thresholding so we only show results for DS in the plots.

In the simulations, we set $N = 10000$ and generate signals and observations according to the model in Section \ref{sec:prob}.  Except where indicated, the signal mean $\mu_{0}$ is normalized to $1$ and the signal standard deviation $\sigma_{0}$ is set to $1/4$.  The identity effort function $h(\lambda) = \lambda$ is used throughout.

%
%
%
%

Two families of generalized OLFC policies are considered, one optimized for MSE (final exponent $\gamma^{(T)}(T-1) = 1/2$, denoted OLFC-MSE) and the other for MAE ($\gamma^{(T)}(T-1) = 2/3$, denoted OLFC-MAE).  The number of stages $T$ is varied from $2$ to $10$ and the final estimate is given by $\mbmu(T)$.  In the offline determination of the parameters $\beta^{(T)}(t)$ and $\gamma^{(T)}(t)$, the optimization in \eqref{eqn:beta0Nested} may be inexact because of finite-sample approximations to the expectations.  To mitigate such errors, we make use of the empirical observation that $\beta^{(T)}(0)$ and $\gamma^{(T)}(0)$ appear to vary smoothly with SNR, and $\beta^{(T)}(0)$ also appears to decrease monotonically with $T$.  Accordingly, we first obtain raw estimates of $\beta^{(T)}(0)$ and $\gamma^{(T)}(0)$ 
and then perform a polynomial fit as a function of SNR, where the polynomials for $\beta^{(T)}(0)$ are constrained to satisfy $\beta^{(T)}(0) \geq \beta^{(T+1)}(0)$ for all $T$.  In our experience, a polynomial degree of $6$ is sufficient to capture the variation of the parameters over the SNR range considered.  

For the rollout OLFC policies, the fractions $\betat^{(T)}(t)$ in \eqref{eqn:betaRollout} are also determined through finite-sample approximations to expectations and are thus subject to the same type of error.  The difference as noted in Section \ref{subsec:policyOLFCRollout} is that \eqref{eqn:betaRollout} must be evaluated online, and hence the number of samples is limited by computational constraints.  To circumvent this tradeoff, we again make use of an empirical smoothness property, this time of the expectation in \eqref{eqn:betaRollout} as a function of $\beta(t)$.  Approximations to the expectations are first obtained using a relatively small number of samples, and a fourth-order polynomial in $\beta(t)$ is fit to the approximation.  The polynomial fit is then minimized to determine $\betat^{(T)}(t)$.  Note that $\beta(t) = 1$ corresponds to a single-stage policy whose cost can be computed exactly from the current state $\mbx(t)$ as described in Appendix \ref{app:OLC}.  Thus $\beta(t) = 1$ and its corresponding single-stage cost represent a fixed point that constrains the polynomial fit.


For DS, while \cite{haupt2011} prescribes a single value for $T$ as a function of the dimension $N$, in our simulations we consider all values of $T$ between $2$ and $10$ as with OLFC.  Following \cite{haupt2011}, we use a geometrically decreasing allocation of effort over stages with decay ratio $3/4$ and equal first and last stages.  More precisely, defining $\alpha(t)$ as the fraction of the total budget used in stage $t$, we have $\alpha(t) = \alpha(0) (3/4)^{t}$ for $t = 1,\ldots,T-2$, $\alpha(T-1) = \alpha(0)$, and $\alpha(0)$ chosen such that $\sum_{t=0}^{T-1} \alpha(t) = 1$.  
%
%

\begin{figure*}[t]
\centerline{
\subfigure[]{\includegraphics[width=0.5\textwidth]{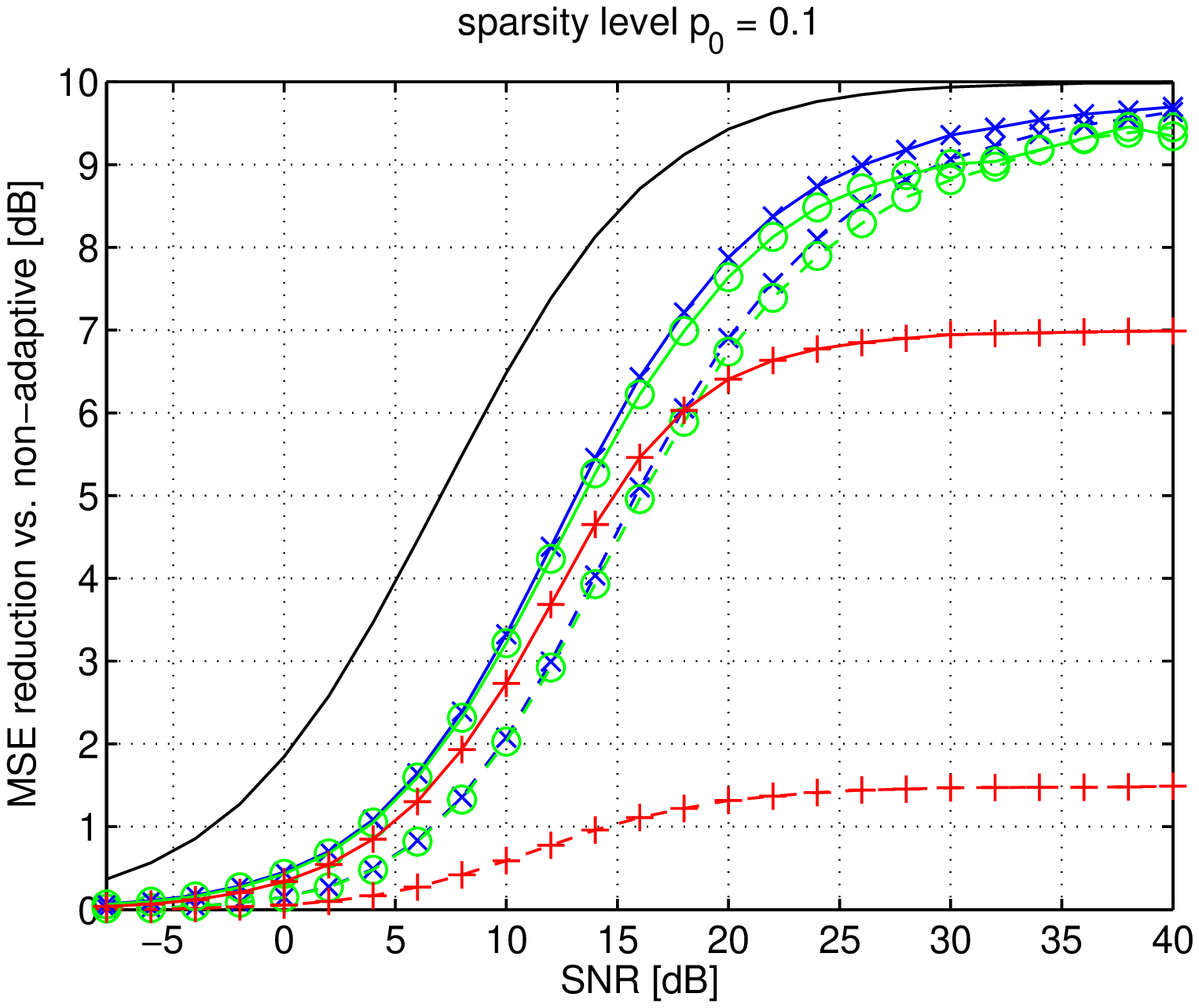}
\label{fig:gainSNRc1p1}}
\subfigure[]{\includegraphics[width=0.5\textwidth]{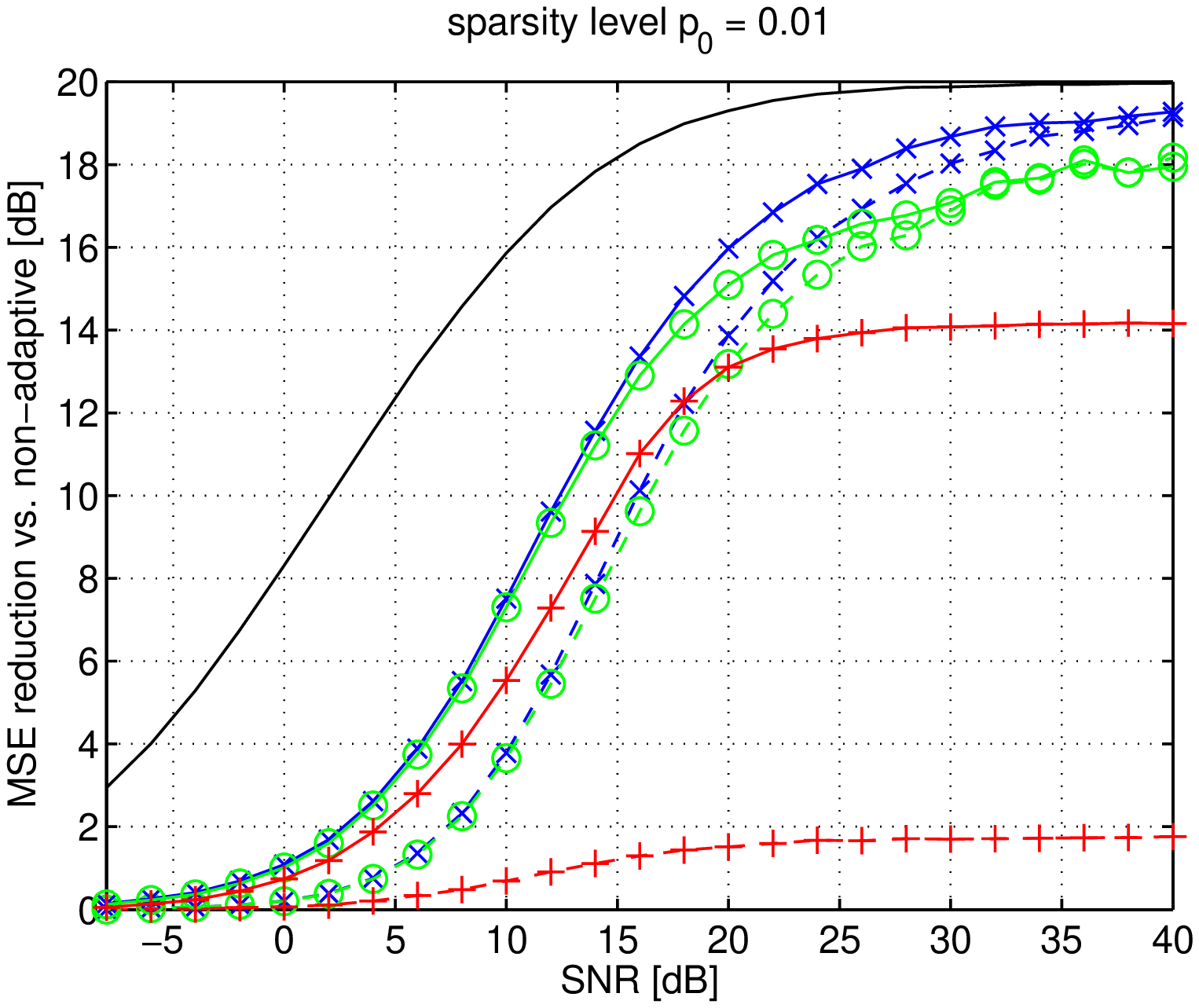}
\label{fig:gainSNRc1p2}}}
\centerline{
\subfigure[]{\includegraphics[width=0.5\textwidth]{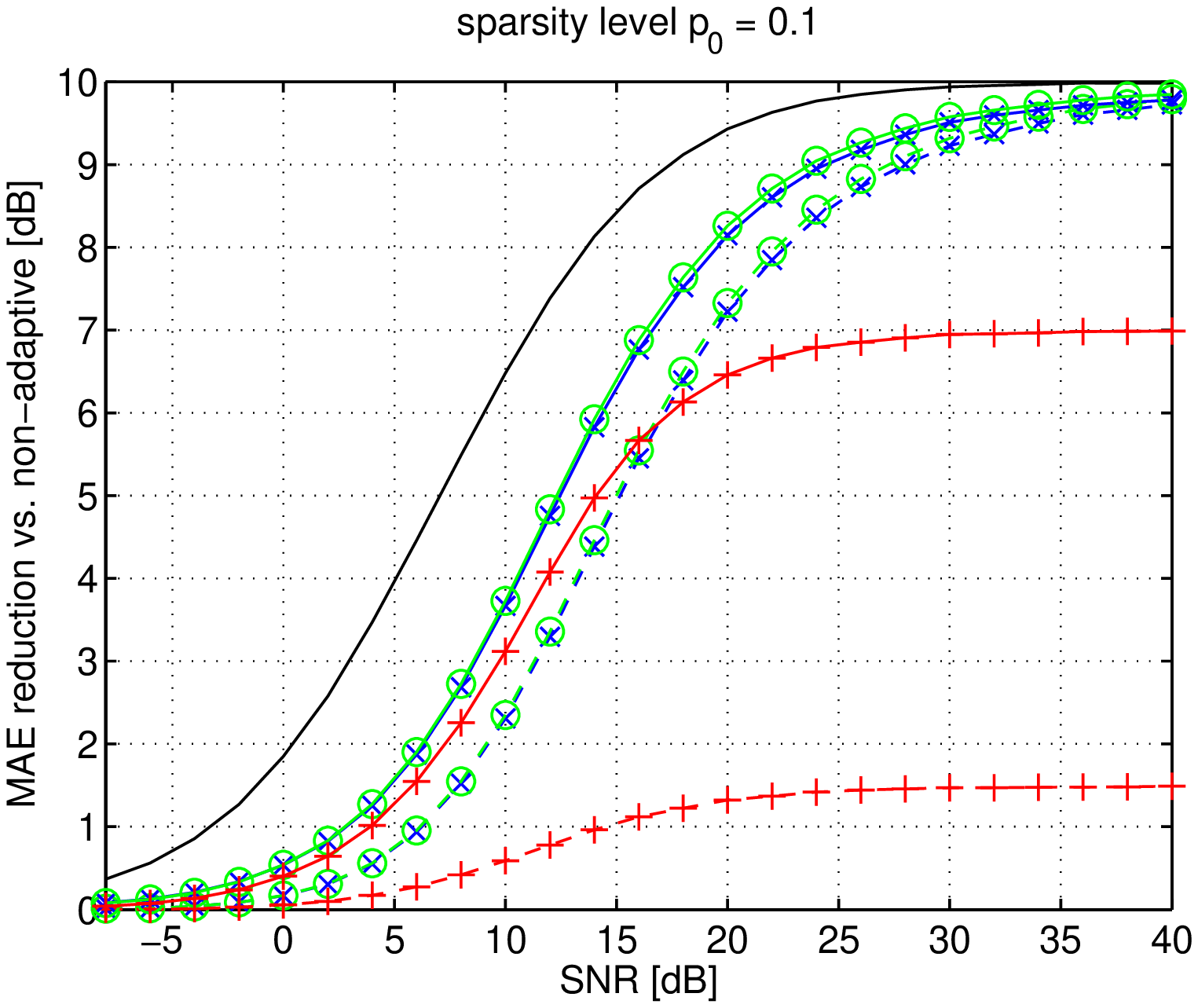}
\label{fig:gainSNRc2p1}}
\subfigure[]{\includegraphics[width=0.5\textwidth]{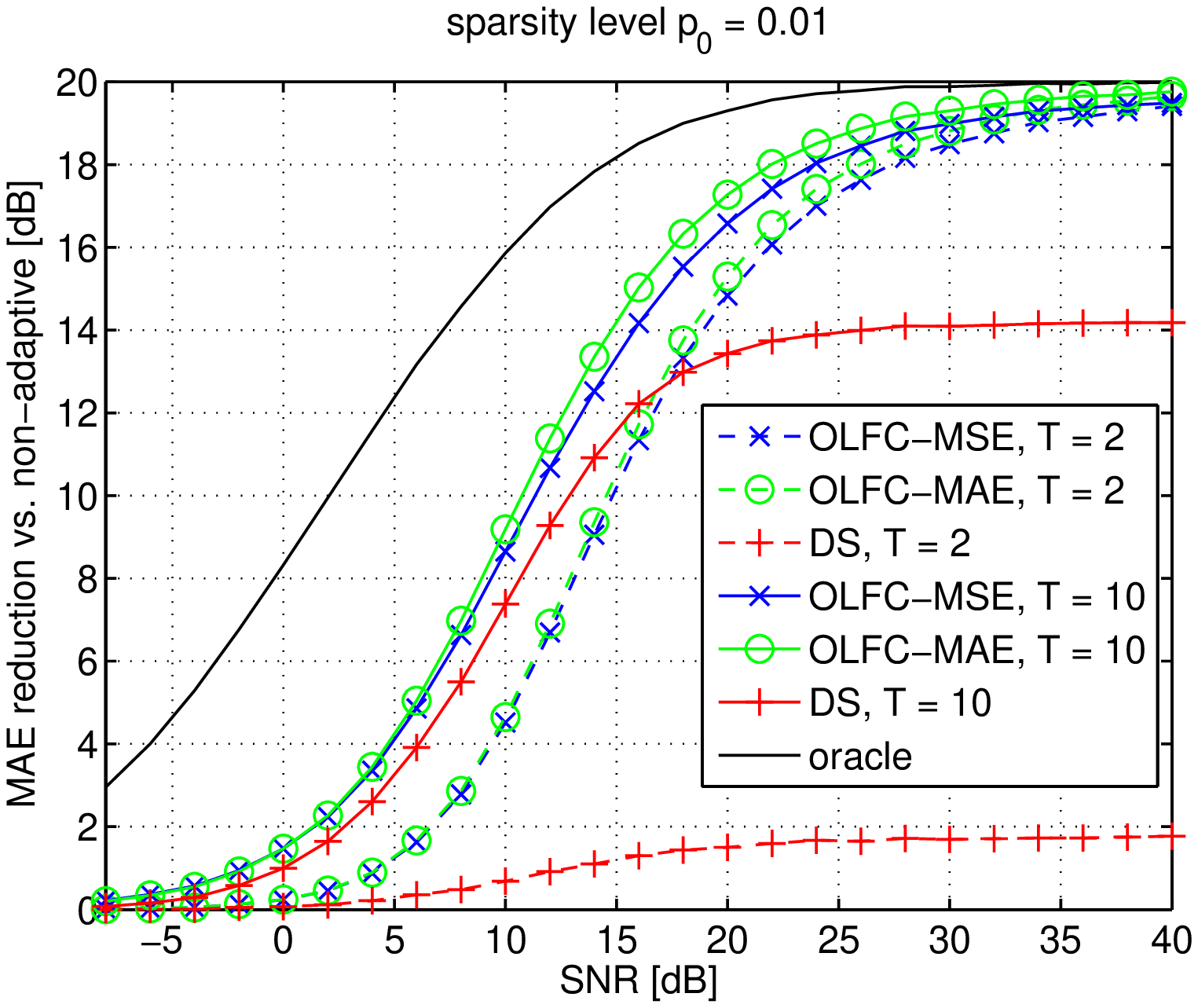}
\label{fig:gainSNRc2p2}}}
\caption{Reduction in MSE (first row) and MAE (second row) relative to non-adaptive estimation as a function of SNR.  The $10$-stage generalized open-loop feedback control (OLFC) policies improve upon the $2$-stage OLFC policies with maximum gains of $1.5$ dB for $p_{0} = 0.1$ and $4.5$ dB for $p_{0} = 0.01$.  The $2$-stage OLFC policies are optimal for $T = 2$.  As the SNR increases, the proposed OLFC policies approach the oracle gain of $1/p_{0}$ and outperform distilled sensing (DS) by several dB.}
\label{fig:gainSNR}
\end{figure*}

In Fig.~\ref{fig:gainSNR}, we plot the MSE (\eqref{eqn:expLossROI} with $L(a) = a^{2}$) and MAE (\eqref{eqn:expLossROI} with $L(a) = a$) for various policies as a function of SNR, where SNR is defined as $10 \log_{10}(\mu_{0}^{2} / \sigma^{2})$ in dB.  Each point represents the average of $4000$ simulations.  The baseline corresponding to $0$ dB on the vertical axis is the optimal non-adaptive policy, which under a uniform prior allocates one unit of effort to all components.  For context, we also include the oracle policy, which distributes effort uniformly over the true signal support.  The oracle thus provides an upper bound on the achievable performance, although the bound is unlikely to be tight at lower SNR.  

In general, adaptivity yields higher gains for sparser signals $(p_{0} = 0.01$) since resources can be concentrated on fewer components once the support is identified.  The $10$-stage generalized OLFC policies improve upon the $2$-stage OLFC policies as expected.  The largest gains occur at intermediate SNR and reach $1.5$ dB for $p_{0} = 0.1$ and $4.5$ dB for $p_{0} = 0.01$.  Recall that the $2$-stage OLFC-MSE policy is optimal in terms of MSE for $T = 2$, and similarly for OLFC-MAE.  Note also that the performance is only slightly affected by a mismatch between the OLFC policy and the loss function.  At high SNR, the OLFC policies approach the oracle gain, which in turn approaches the sparsity factor $1/p_{0}$.  In contrast, the DS policies saturate at significantly lower levels since they are not designed with estimation performance in mind.  While the $10$-stage DS policy outperforms the optimal $2$-stage policy at lower SNR, the $10$-stage OLFC policies have the best performance at all SNR.  


\begin{figure}[t]
\centering
\ifCLASSOPTIONdraftcls
\includegraphics[width=0.7\columnwidth]{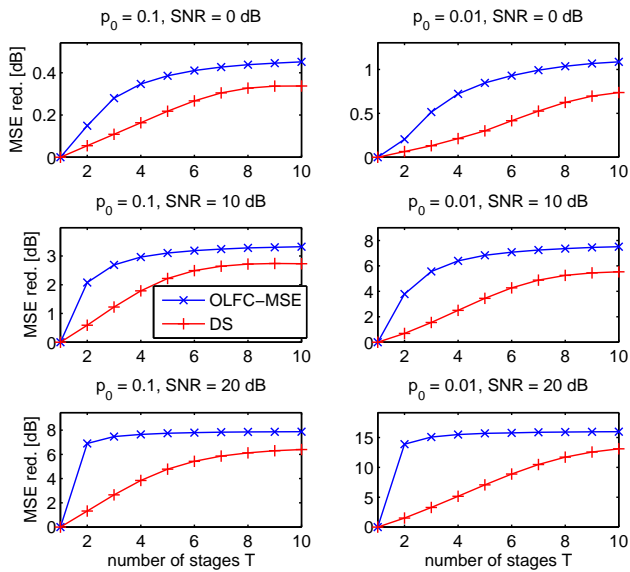}
\else
\includegraphics[width=0.95\columnwidth]{gainT.eps}
\fi
\caption{MSE reduction as a function of the number of stages $T$.  Gains diminish as $T$ increases but less quickly at lower SNR and higher sparsity.  In all cases shown, the proposed OLFC-MSE policy with $5$ stages performs better than a $10$-stage DS policy.}
\label{fig:gainT}
\end{figure}

Fig.~\ref{fig:gainT} shows decreases in MSE with the number of stages $T$.  The incremental gains predicted by Proposition \ref{prop:nested} diminish as $T$ increases.  Using more stages is more beneficial at lower SNR and higher sparsity, whereas at higher SNR most of the signal components can be located in a single step and a two-stage OLFC policy performs almost as well as a policy with many more stages.  The gains for DS do not diminish as quickly but are lower overall, never exceeding the gain of the corresponding $5$-stage generalized OLFC policy.

\begin{figure*}[t]
\centerline{
\subfigure[]{\includegraphics[width=0.5\textwidth]{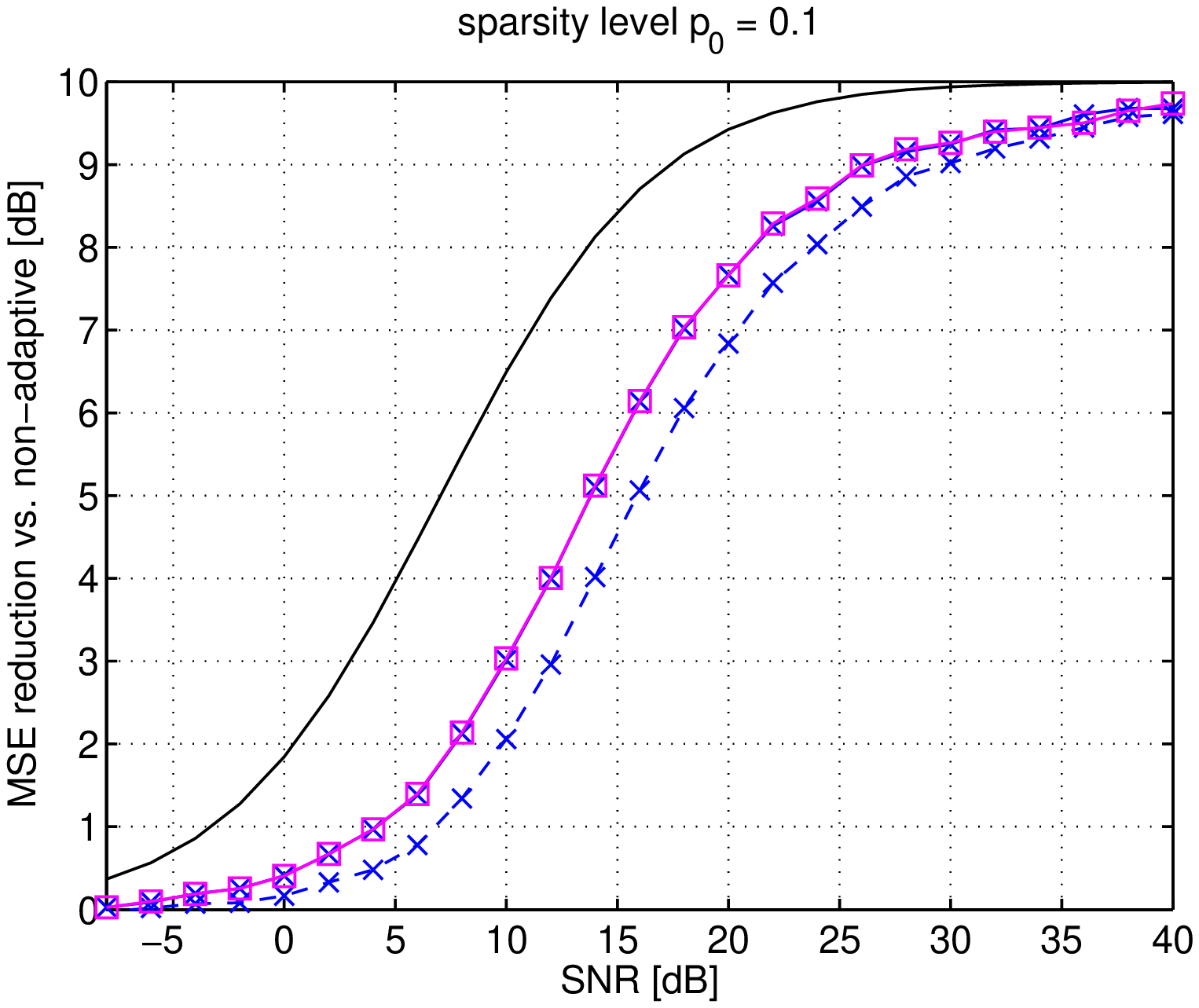}
\label{fig:gainSNRrollp1}}
\subfigure[]{\includegraphics[width=0.5\textwidth]{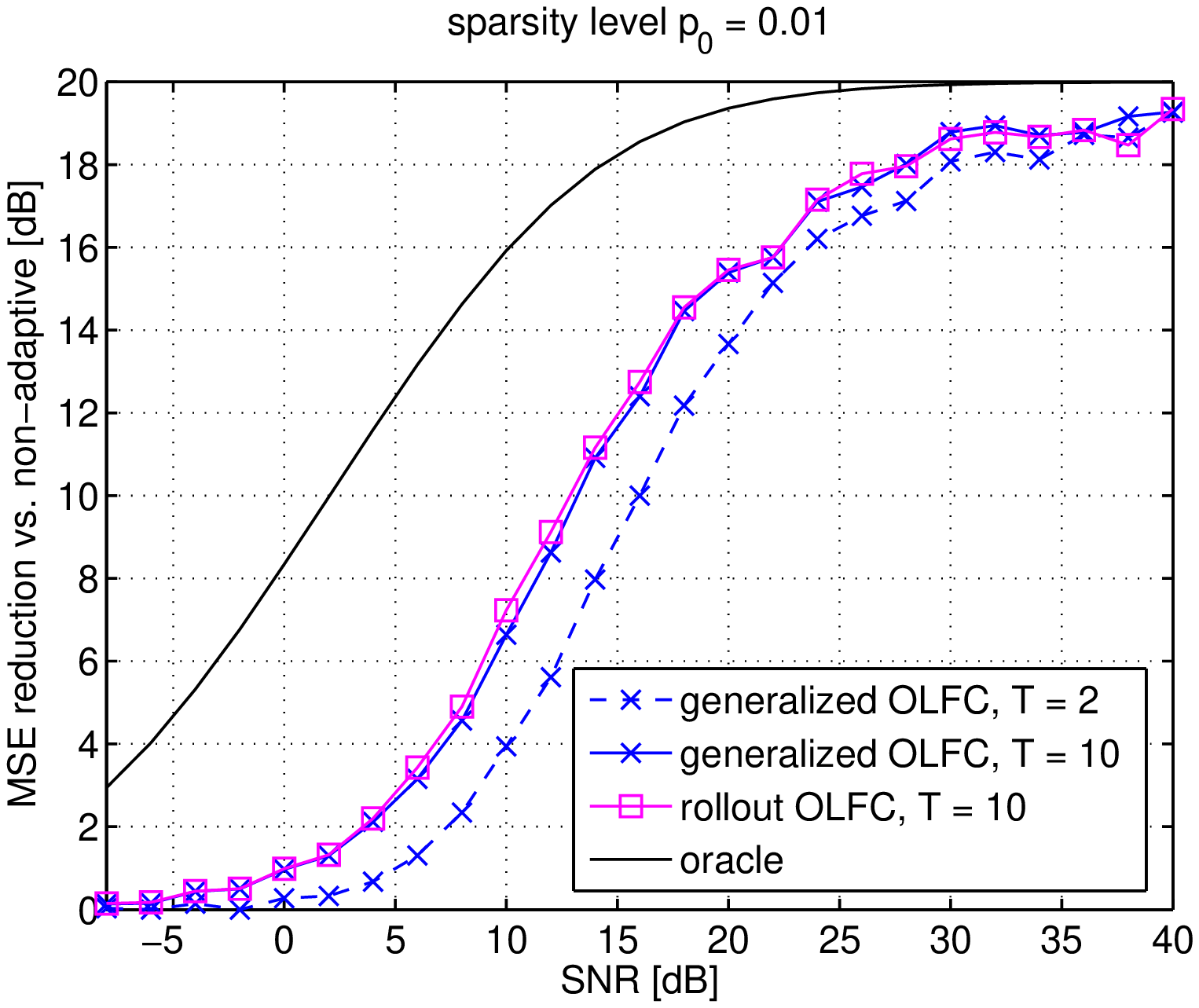}
\label{fig:gainSNRrollp2}}}
\caption{Comparison of generalized and rollout OLFC policies as a function of SNR.  The improvement due to rollout is minimal.}
\label{fig:gainSNRroll}
\end{figure*}

In Fig.~\ref{fig:gainSNRroll}, we consider the performance improvement due to policy rollout, as guaranteed by Proposition \ref{prop:rollout}.  For this experiment only, the exponent $\gamma$ in \eqref{eqn:orderingOLFC}--\eqref{eqn:C} is fixed at $2/(q+2) = 1/2$ ($q=2$ for MSE).  The dimension $N$ is lowered to $1000$ and the results are averaged over only $1000$ simulations because of the higher computational complexity of rollout.  For $p_{0} = 0.1$ in Fig.~\ref{fig:gainSNRrollp1}, no decrease in MSE is seen, whereas for $p_{0} = 0.01$ in Fig.~\ref{fig:gainSNRrollp2}, the decrease is never more than $0.6$ dB.  It appears therefore that for the problem at hand, the performance gained from rollout is minimal while the computational cost of the required online simulations is much greater. 

\begin{figure*}[t]
\centerline{
\subfigure[]{\includegraphics[width=0.5\textwidth]{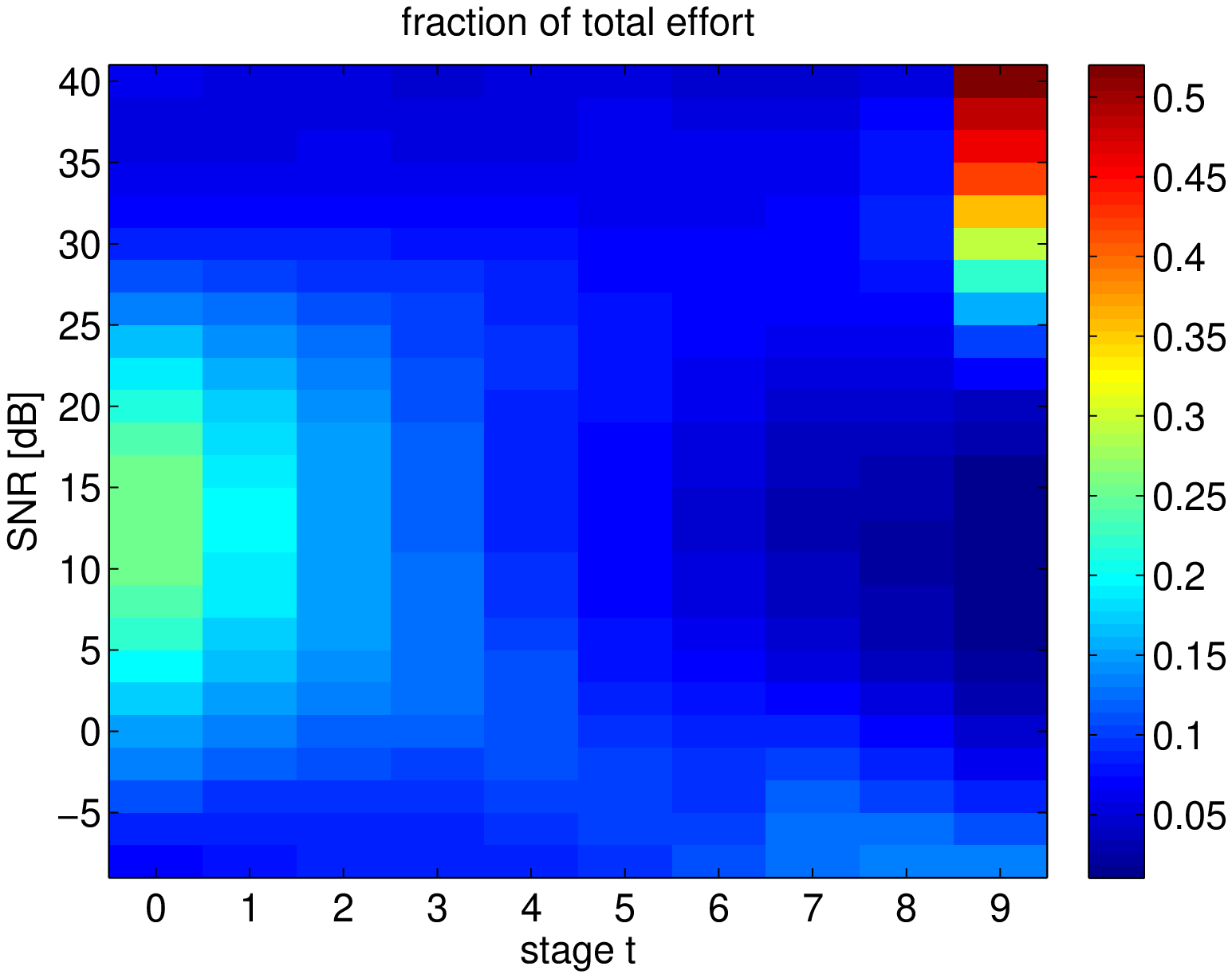}
\label{fig:alphatSNR}}
\subfigure[]{\includegraphics[width=0.5\textwidth]{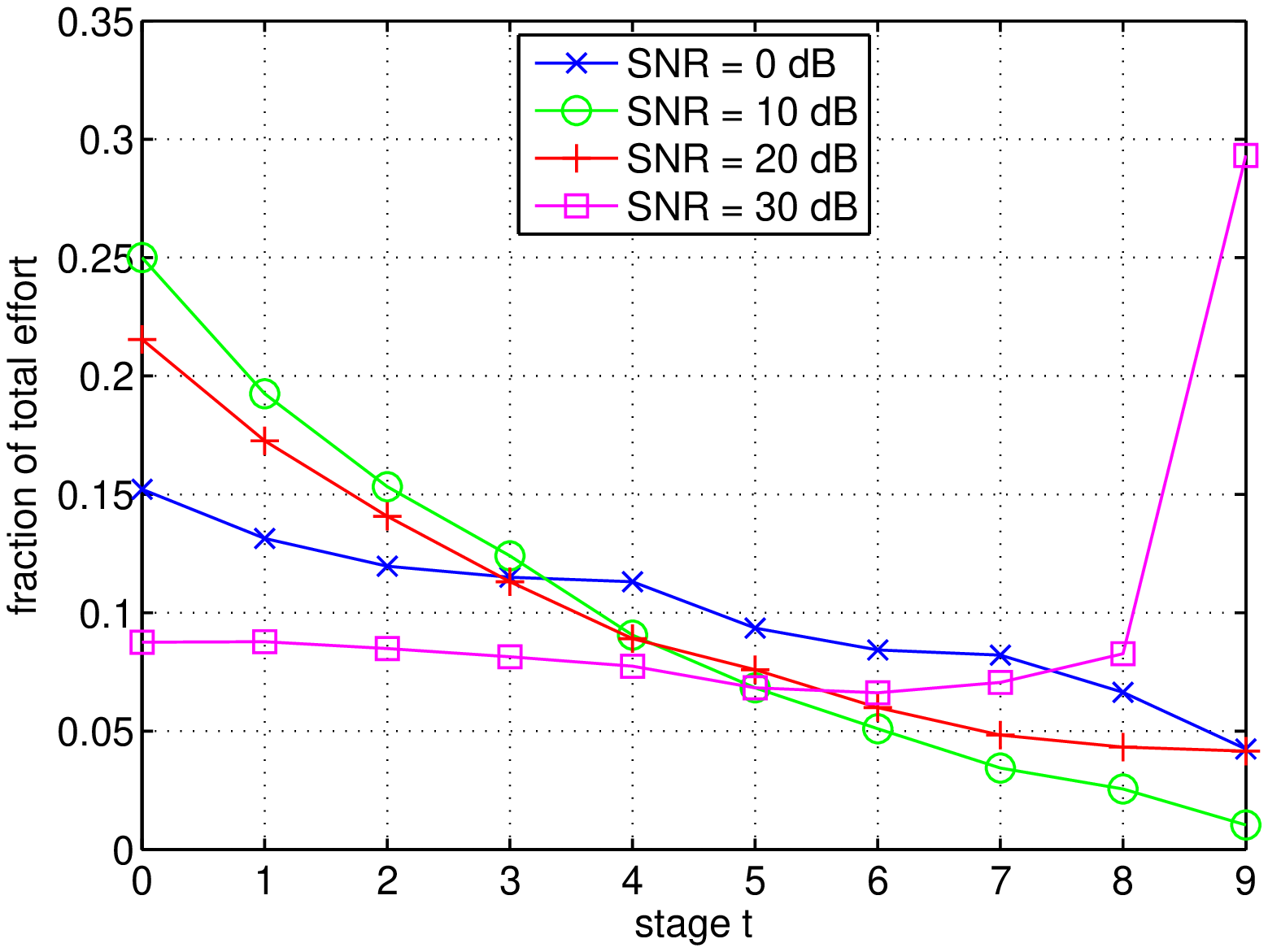}
\label{fig:alphat}}}
\caption{Fraction of total budget allocated to each stage in a $10$-stage OLFC-MSE policy for $p_{0} = 0.01$ and (a) all SNR values, (b) SNR $= 0$, $10$, $20$, $30$ dB.  Three regimes can be seen in (a): a nearly uniform regime below $0$ dB SNR, a decaying ``distilling'' regime between $0$ and $25$ dB, and a near-oracle regime above $25$ dB with increasing emphasis on the last stage.}
\label{fig:alpha}
\end{figure*}

Fig.~\ref{fig:alpha} depicts the fraction $\alpha^{(T)}(t)$ of the total budget allocated to each stage in a $10$-stage OLFC-MSE policy for different SNR levels and $p_{0} = 0.01$.  The fractions $\alpha^{(T)}(t)$ are related to the fractions $\beta^{(T)}(t)$ of the remaining budget through a straightforward transformation.
%
%
Three regimes may be distinguished in Fig.~\ref{fig:alphatSNR}. 
At very low SNR, it is difficult to identify the signal support and the allocation is close to uniform. 
Between $0$ and $25$ dB SNR, the allocation is heavily weighted toward earlier stages.  As seen in Fig.~\ref{fig:alphat}, the decrease with time is reminiscent of the geometric decay prescribed by distilled sensing.  Above $25$ dB SNR, the support can be determined with relatively little effort and an increasing fraction of the budget is reserved for the last stage to exploit this knowledge.


The proposed policies are based on a Bayesian framework and are thus dependent on prior knowledge of the expected sparsity level and SNR, specifically in the form of the parameters $p_{0}$, $\mu_{0}$, and $\sigma^{2}$.  If these prior parameters are misspecified, the correct values can be learned through the Bayesian update process \eqref{eqn:stateEvol} but some degradation in performance is to be expected.  One possible remedy is to introduce hyper-parameters for $p_{0}$, $\mu_{0}$, and $\sigma^{2}$, but this approach is more complicated and is beyond the scope of the current paper.  Moreover, as will be seen shortly, the effect of mismatched priors on the generalized OLFC policies is quite mild except when the SNR is overestimated.

\begin{figure*}[t]
\centerline{
\subfigure[]{\includegraphics[width=0.33\textwidth]{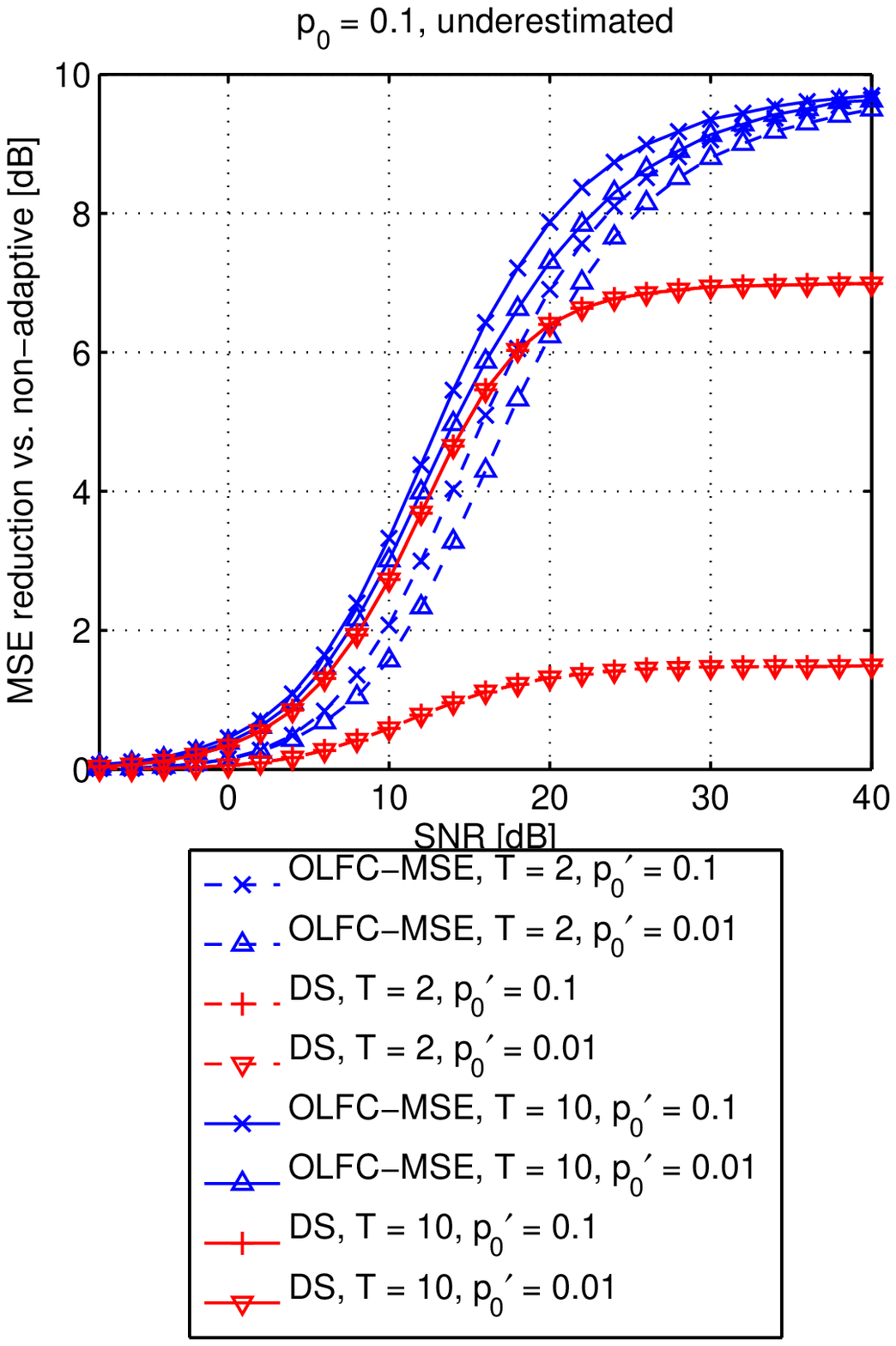}
\label{fig:misp1}}
\subfigure[]{\includegraphics[width=0.33\textwidth]{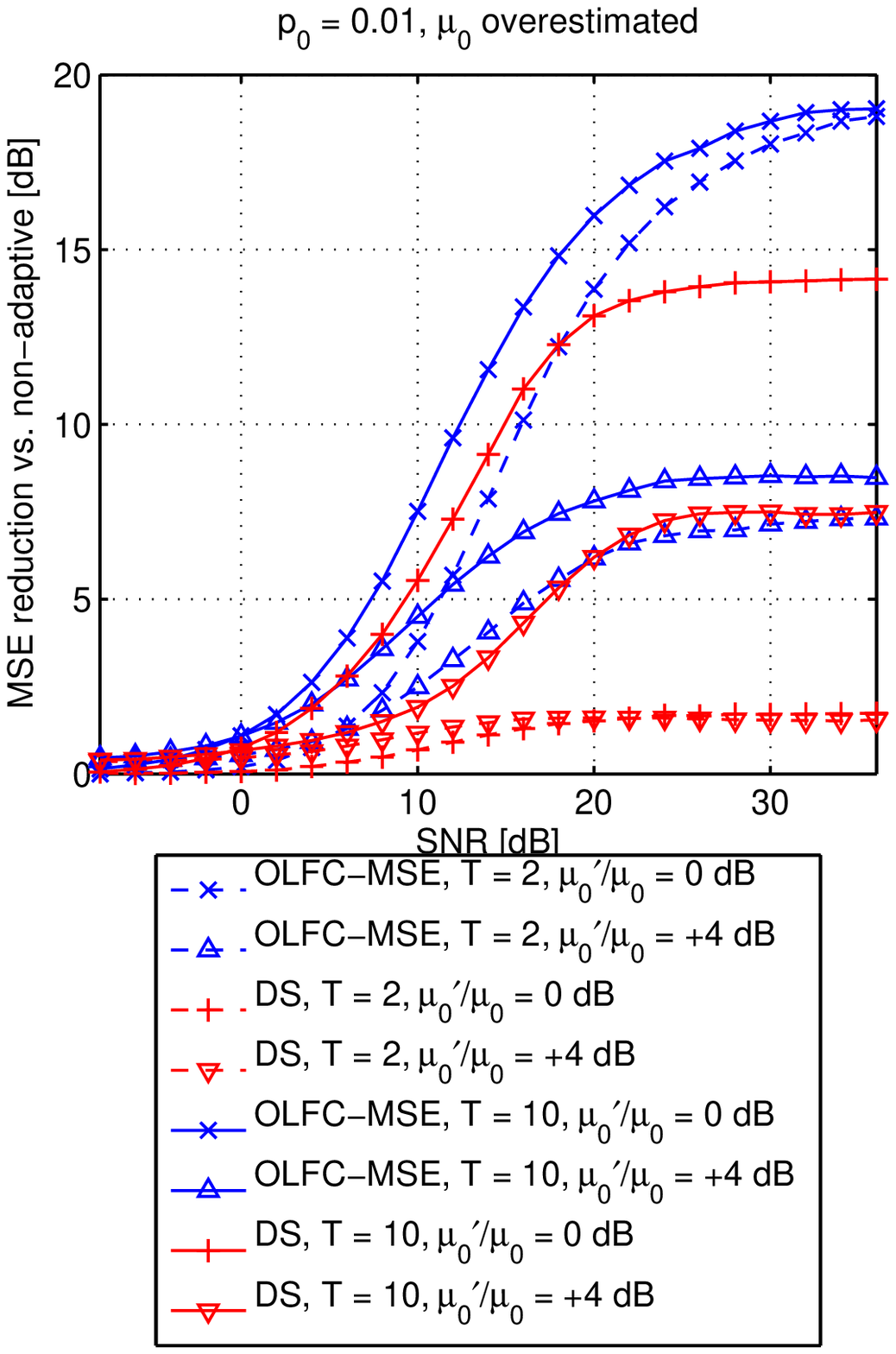}
\label{fig:mismuop2}}
\subfigure[]{\includegraphics[width=0.33\textwidth]{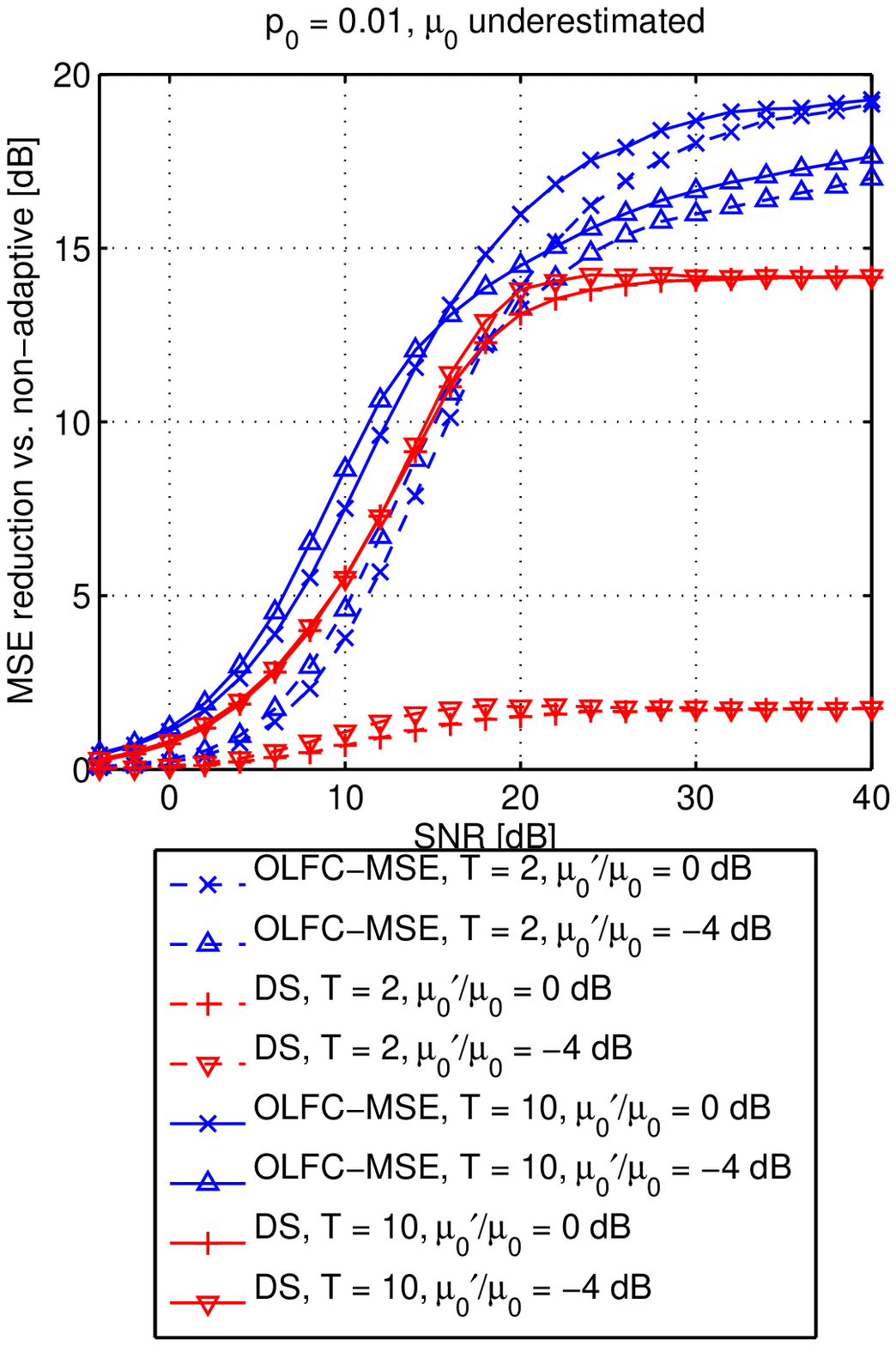}
\label{fig:mismuup2}}}
\caption{Reduction in MSE relative to non-adaptive estimation as a function of SNR under mismatches in prior parameters.  In (a), $p_{0}$ is underestimated by an order of magnitude and the effects are minor.  More severe losses are seen in (b) with a $4$ dB overestimate of $\mu_{0}$.  In (c), $\mu_{0}$ is underestimated by $4$ dB and the losses are again modest.  In all cases, the proposed OLFC-MSE policy remains better than DS.}
\label{fig:mis}
\end{figure*}

To assess the effect of mismatched priors on the generalized OLFC policies, a series of experiments are conducted in which one of $p_{0}$, $\mu_{0}$, or $\sigma^{2}$ is misspecified.  In Fig.~\ref{fig:misp1}, the true sparsity level $p_{0}$ is $0.1$ while the value $p'_{0}$ assumed by the policies is either $0.1$ or $0.01$.  The performance loss of the OLFC-MSE policies is rather mild given the order-of-magnitude underestimate of $p_{0}$.  Similar results are seen when $p_{0}$ is overestimated.  DS on the other hand does not make use of the parameter $p'_{0}$ and is therefore unaffected.  

In Figs.~\ref{fig:mismuop2} and \ref{fig:mismuup2}, $p_{0}$ is set to $0.01$ while the signal mean $\mu'_{0}$ assumed by the policies is either correct or off by $\pm 4$ dB.  The signal standard deviation $\sigma_{0}$ is also changed to $0.40$, making the mean mismatches on the order of one standard deviation.  As can be seen from \eqref{eqn:muEvol}, a misspecification of $\mu_{0}$ leads to a biased estimate $\mbmu(T)$, although the bias can be reduced by allocating more effort to the observations.  It is clear from Figs.~\ref{fig:mismuop2}(c) that overestimating $\mu_{0}$ results in more significant losses due to missed detections of weaker than expected signal components, especially at high SNR.  In contrast, when $\mu_{0}$ is underestimated, the reduction in MSE relative to nonadaptive sampling can actually be greater than in the matched case; this can be attributed to a reduction in bias.  In both cases, the OLFC-MSE policy remains better than DS.  The consequences of misspecifying $\sigma^{2}$ are less severe than for $\mu_{0}$ with underestimating $\sigma^{2}$ being worse.  These findings suggest that the policies are more sensitive to overestimates of the SNR than underestimates.

\section{Application to radar imaging}
\label{sec:SARtanks}

\begin{figure}[t]
\centerline{
\subfigure[]{\includegraphics[width=0.51\columnwidth]{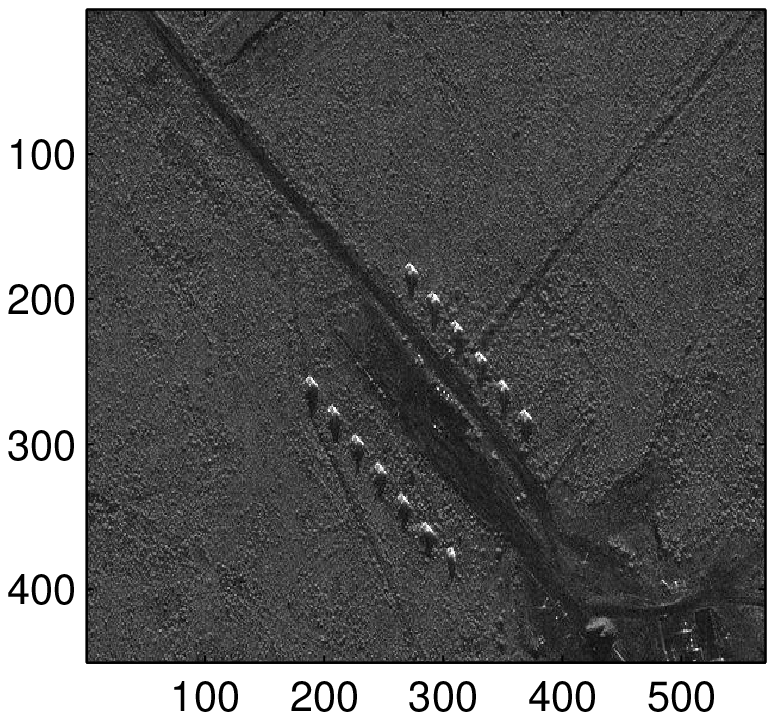}
\label{fig:tanksFull}}
\subfigure[]{\includegraphics[width=0.49\columnwidth]{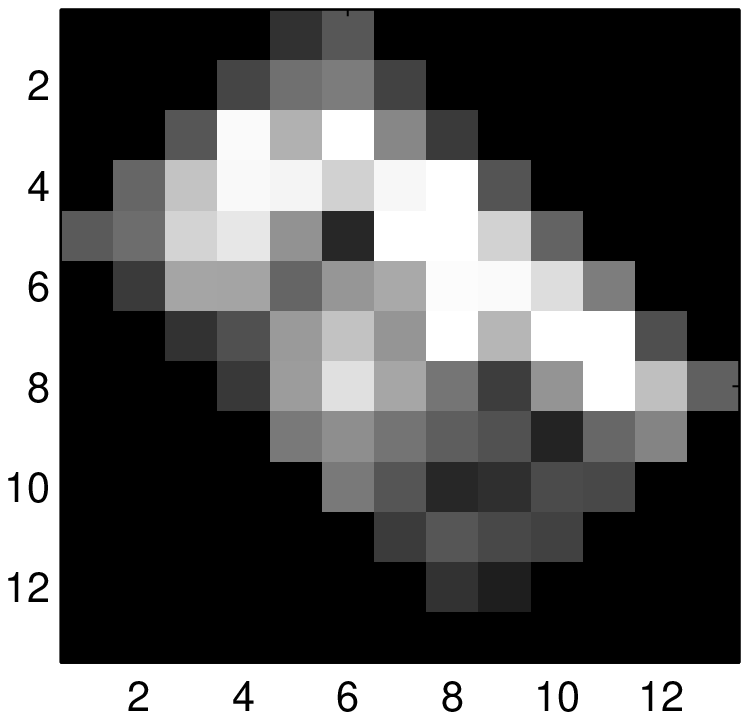}
\label{fig:tank}}
}
\caption{(a) Original SAR image taken from \cite{sandiaTanks} for radar imaging example. (b) Tank template used for 2-D linear filtering.}
\label{fig:tanks0}
\end{figure}

In this section, the proposed allocation policies are applied to a radar imaging example also considered in \cite{bashan2008}.  The original synthetic aperture radar (SAR) image in Fig.~\ref{fig:tanksFull} shows $13$ tanks in a large field and is therefore sparse in terms of targets.  In the adaptive setting, it is assumed that the position and dwell time of the radar beam can be controlled, and our goal is to illustrate the benefits of such adaptivity in acquiring sparse targets. 

We assume a Swerling II target model, commonly used in radar \cite{meikle2008}, in which the observation $z_{i}(t)$ of location $i$ in stage $t$ is given by the empirical mean 
\begin{equation}\label{eqn:obsSwerling}
z_{i}(t) = \frac{1}{\kappa_{i}(t-1)} \sum_{s=1}^{\kappa_{i}(t-1)} z_{is}(t),
\end{equation}
where the $z_{is}(t)$ are i.i.d.~exponential random variables with mean equal to the true target amplitude $x_{i}$ in Fig.~\ref{fig:tanksFull}, and $\kappa_{i}(t-1)$ is the number of radar pulses.  Thus as $\kappa_{i}(t-1)$ increases, the distribution of $z_{i}(t)$ becomes more concentrated around $x_{i}$.  The total budget consists of $NP$ pulses and the average number of pulses per location $P$ is thus equivalent to SNR.

The Swerling observation model presents a test of robustness of the policies to non-Gaussianity.  Results obtained under Gaussian and speckle noise are similar.  In addition, several accommodations are made to better conform to the model in Section \ref{sec:prob}.  Most notably, while the targets in Fig.~\ref{fig:tanksFull} are indeed sparse, they each extend over several pixels and within this extent, their amplitudes are not uniformly different from the background.  To address this non-uniformity, each observed image $\mbz(t)$ is preprocessed with a 2-D linear filter, following the approach in \cite{bashan2008} and using the same approximate tank template as in \cite[Fig.~6]{bashan2008} and reproduced in Fig.~\ref{fig:tank}.  The filtered images $\mby(t)$ display clusters of uniformly brighter intensities at the locations of the tanks and are used as the input to the effort allocation policies.  
We use $p_{0} = 0.001$ as the initial sparsity estimate in the filtered domain.  The other prior parameters $\mu_{0}$, $\sigma_{0}^{2}$, and $\sigma^{2}$ are estimated from the first-stage filtered observation $\mby(1)$.  More specifically, the background mean (generally nonzero) and variance $\sigma^{2}$ are estimated from the $y_{i}(1)$ below the $1-p_{0}$ quantile, while the initial signal mean $\mu_{0}$ and variance $\sigma_{0}^{2}$ are estimated from the $y_{i}(1)$ above the $1-p_{0}$ quantile.  Once the allocation $\mblambda(t)$ has been determined in each stage, it is mapped to a pulse allocation $\mbkappa(t)$ in the original unfiltered domain by convolving $\mblambda(t)$ as an image with the support of the tank template in Fig.~\ref{fig:tank} (a binary image) and normalizing so that $\sum_{i} \kappa_{i}(t) = \sum_{i} \lambda_{i}(t)$.  The allocation $\mbkappa(t)$ is then rounded to satisfy the integer restriction, again while preserving the sum.

\begin{figure}[t]
\centerline{
\subfigure[]{\includegraphics[width=0.5\columnwidth]{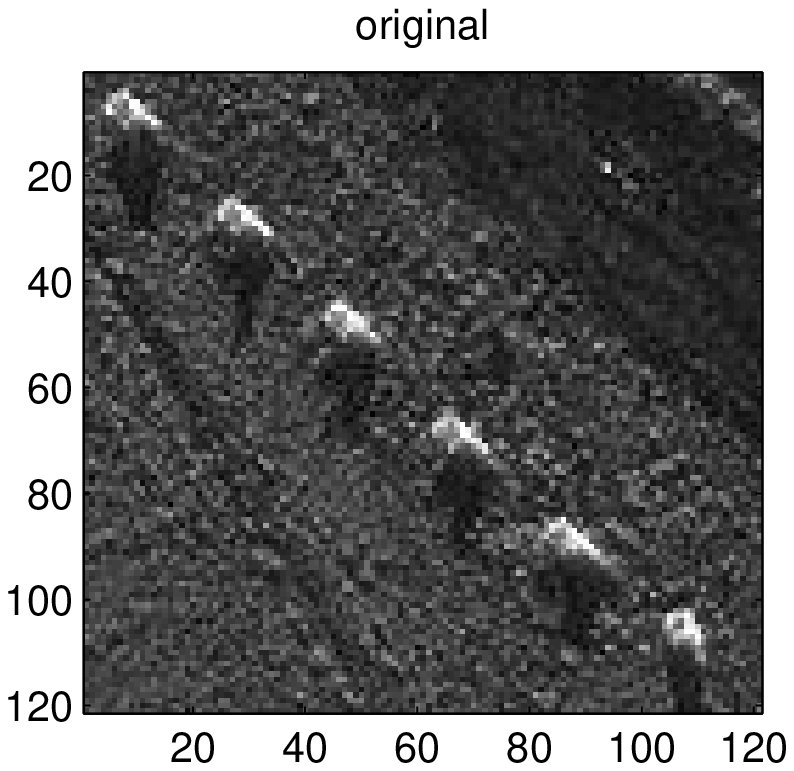}
\label{fig:tanksOrig}}
\subfigure[]{\includegraphics[width=0.5\columnwidth]{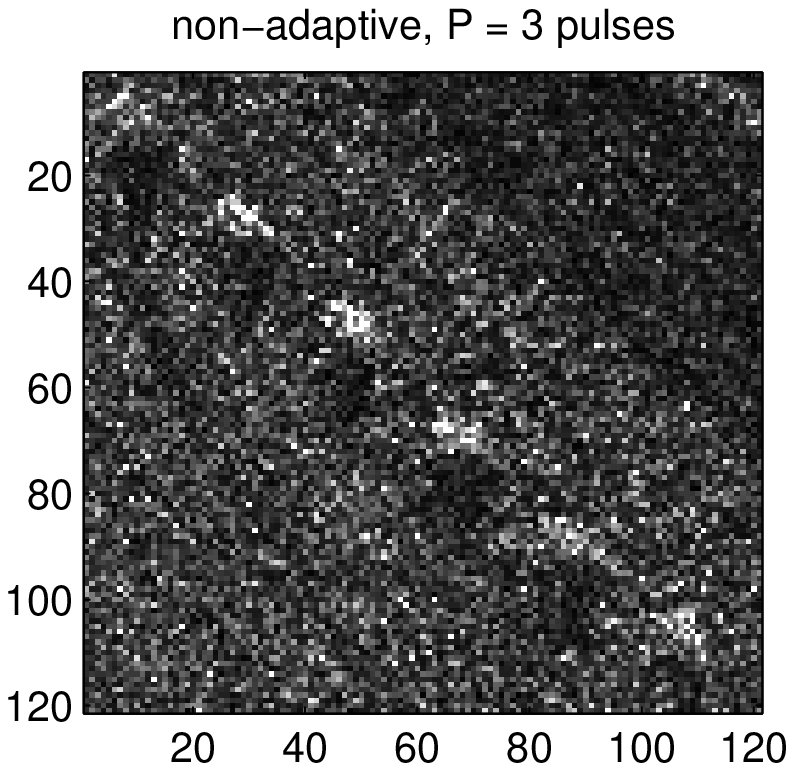}
\label{fig:tanksNonAdap}}
}
\centerline{
\subfigure[]{\includegraphics[width=0.5\columnwidth]{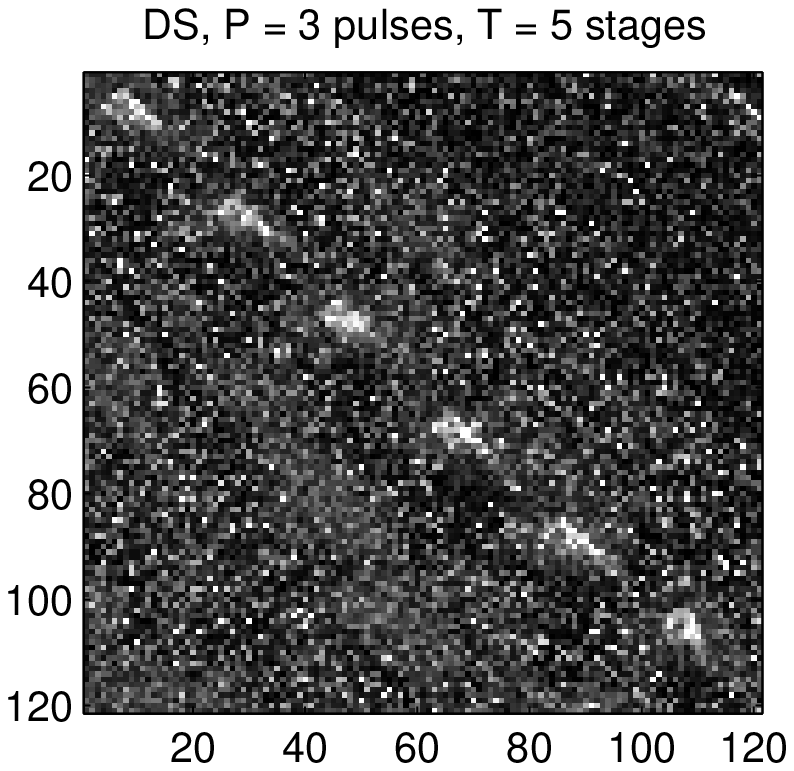}
\label{fig:tanksDS}}
\subfigure[]{\includegraphics[width=0.5\columnwidth]{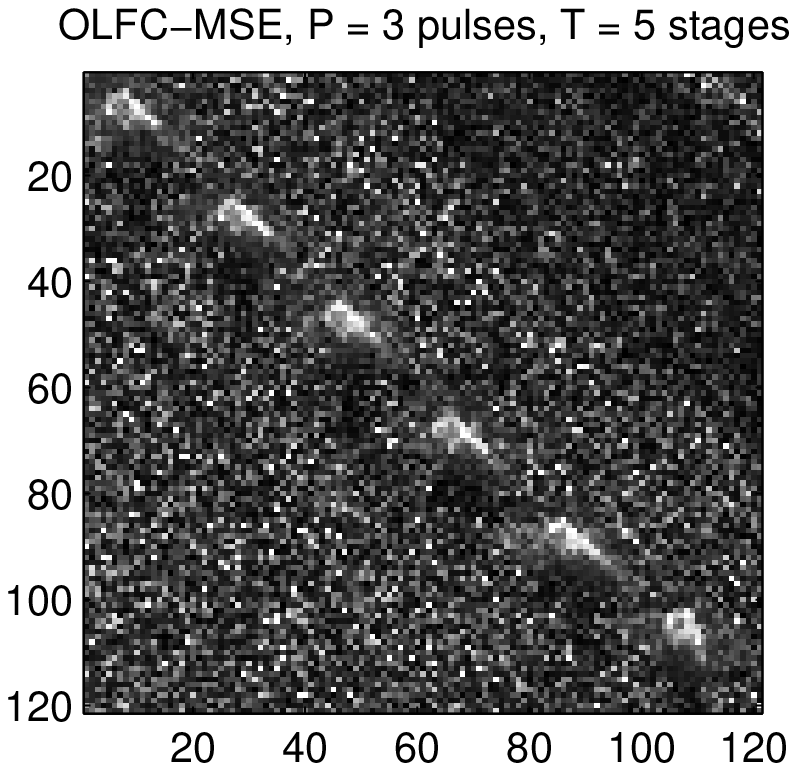}
\label{fig:tanksOLFCMSE}}
}
\caption{Portion of original image (a) in radar imaging example and reconstructions from $P = 3$ pulses per location allocated non-adaptively (b), using $5$-stage DS (c), and using $5$-stage OLFC-MSE (d). OLFC suppresses noise more strongly around each tank and recovers details more faithfully.}
\label{fig:tanks}
\end{figure}

The reconstructed image $\hat{\mbx}$ is formed as a maximum-likelihood estimate of $\mbx$ based on $\mbz(1),\ldots,\mbz(T)$:
\[
\hat{x}_{i} = \frac{\sum_{t=1}^{T} \kappa_{i}(t) z_{i}(t)}{\sum_{t=1}^{T} \kappa_{i}(t)}, \quad i = 1,\ldots,N.
\]
In the non-adaptive single-stage case, this reduces to $\hat{x}_{i} = z_{i}(1)$ with $\kappa_{i}(0) = P$ in \eqref{eqn:obsSwerling}.  Fig.~\ref{fig:tanks} shows a $120\times120$ portion of the original image (the full $450\times570$ image in Fig.~\ref{fig:tanksFull} is used in processing) together with reconstructions from $P = 3$ pulses per location.  We focus attention on the targets of interest, namely the tanks.  In the non-adaptive reconstruction in Fig.~\ref{fig:tanksNonAdap}, the tanks are obscured by noise.  Better images result from the two adaptive policies.  The OLFC reconstruction however shows greater noise suppression around each tank and recovers amplitude details more faithfully.  

\begin{figure}[t]
\centerline{
\subfigure[]{\includegraphics[width=0.5\columnwidth]{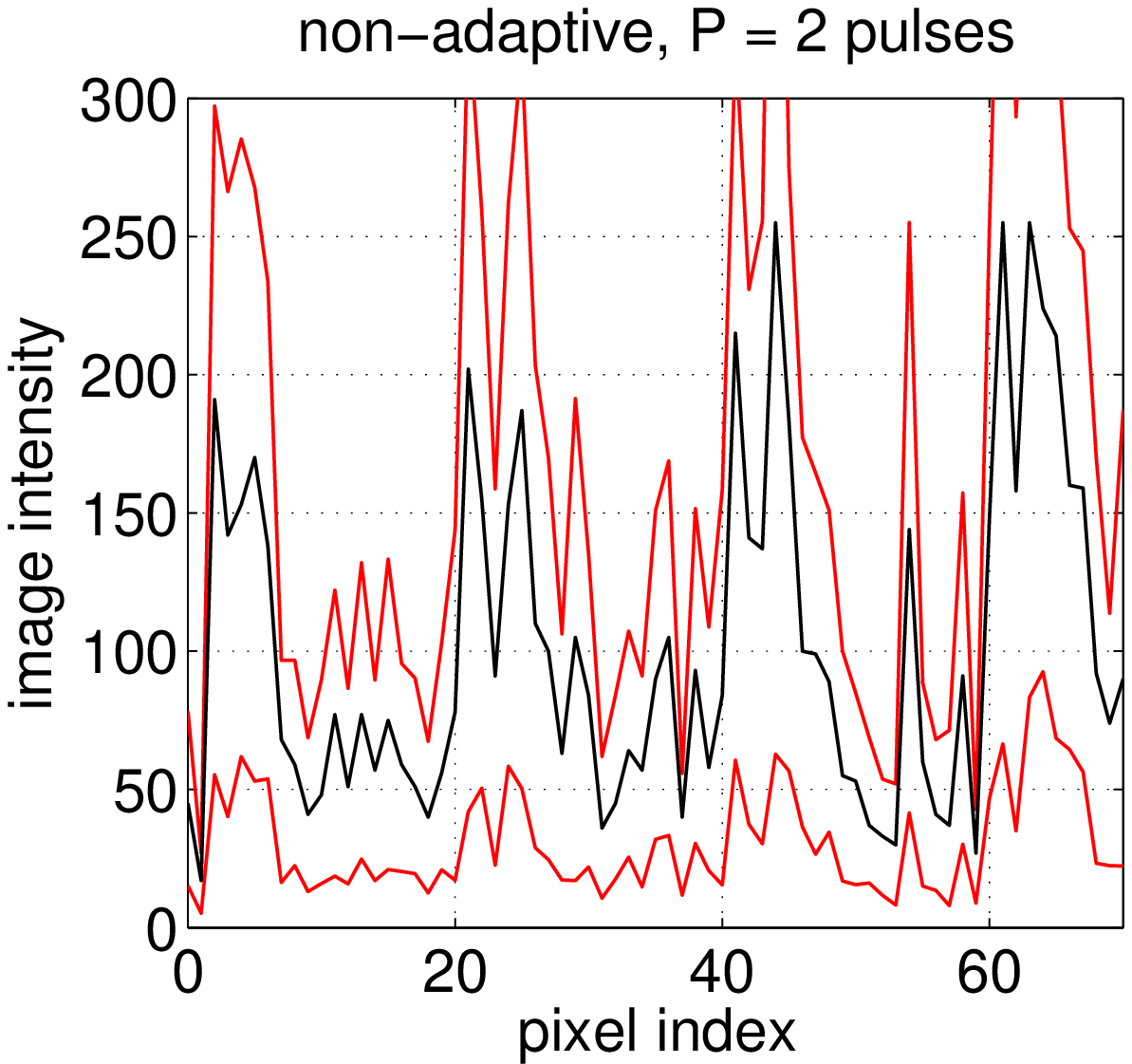}
\label{fig:profileNonAdap}}
\subfigure[]{\includegraphics[width=0.5\columnwidth]{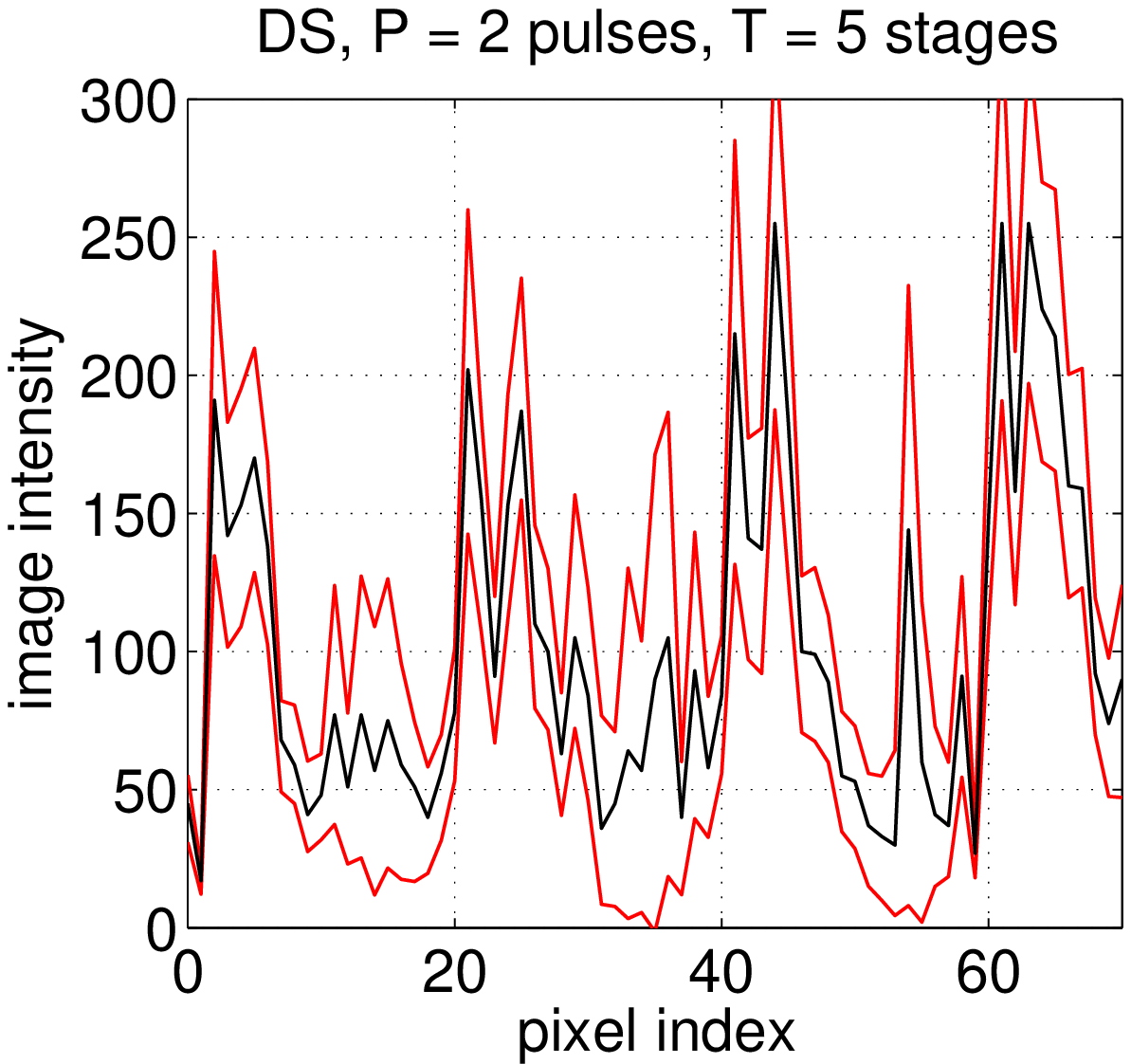}
\label{fig:profileDS}}
}
\centerline{
\subfigure[]{\includegraphics[width=0.5\columnwidth]{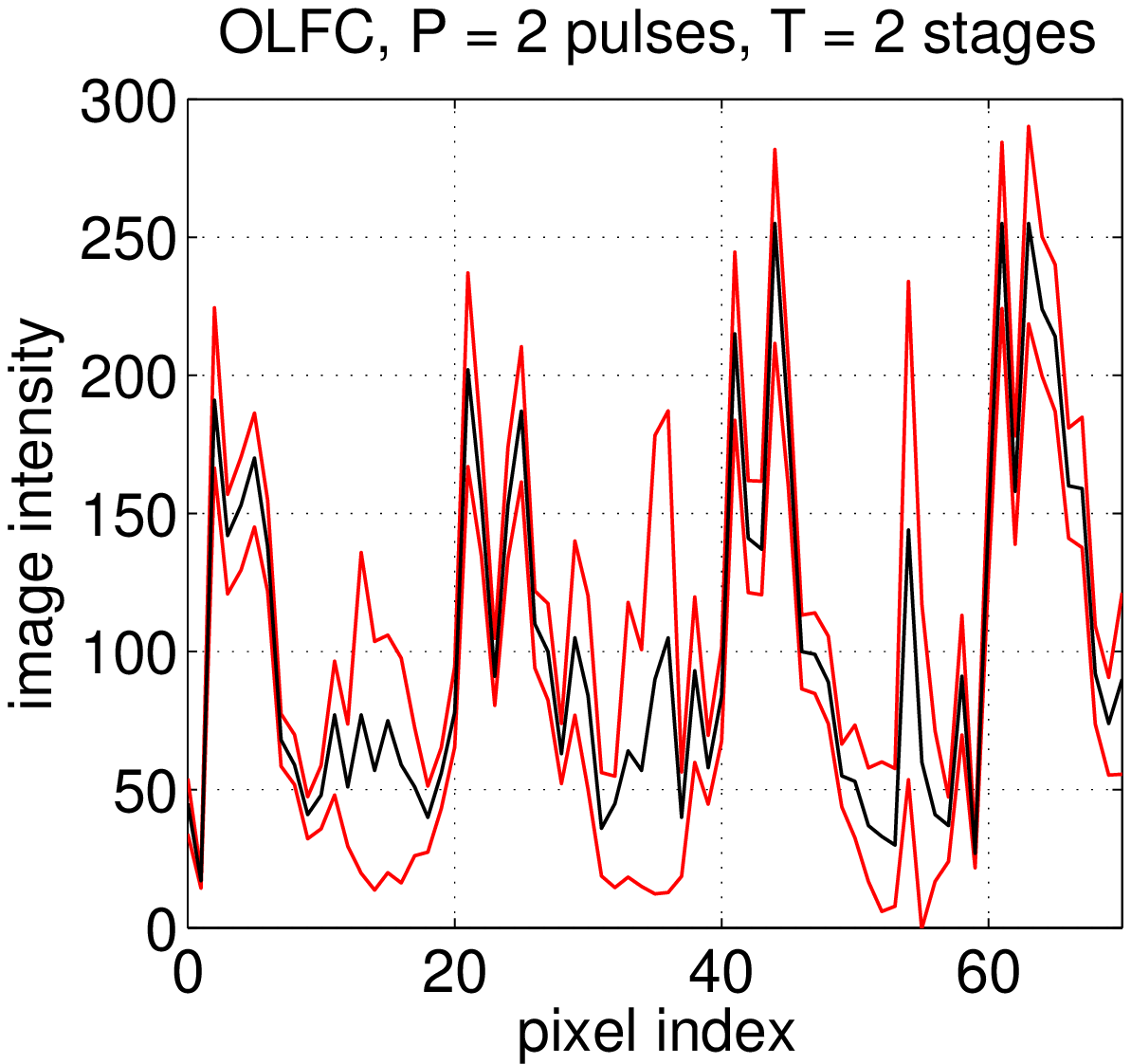}
\label{fig:profileOLFCT2}}
\subfigure[]{\includegraphics[width=0.5\columnwidth]{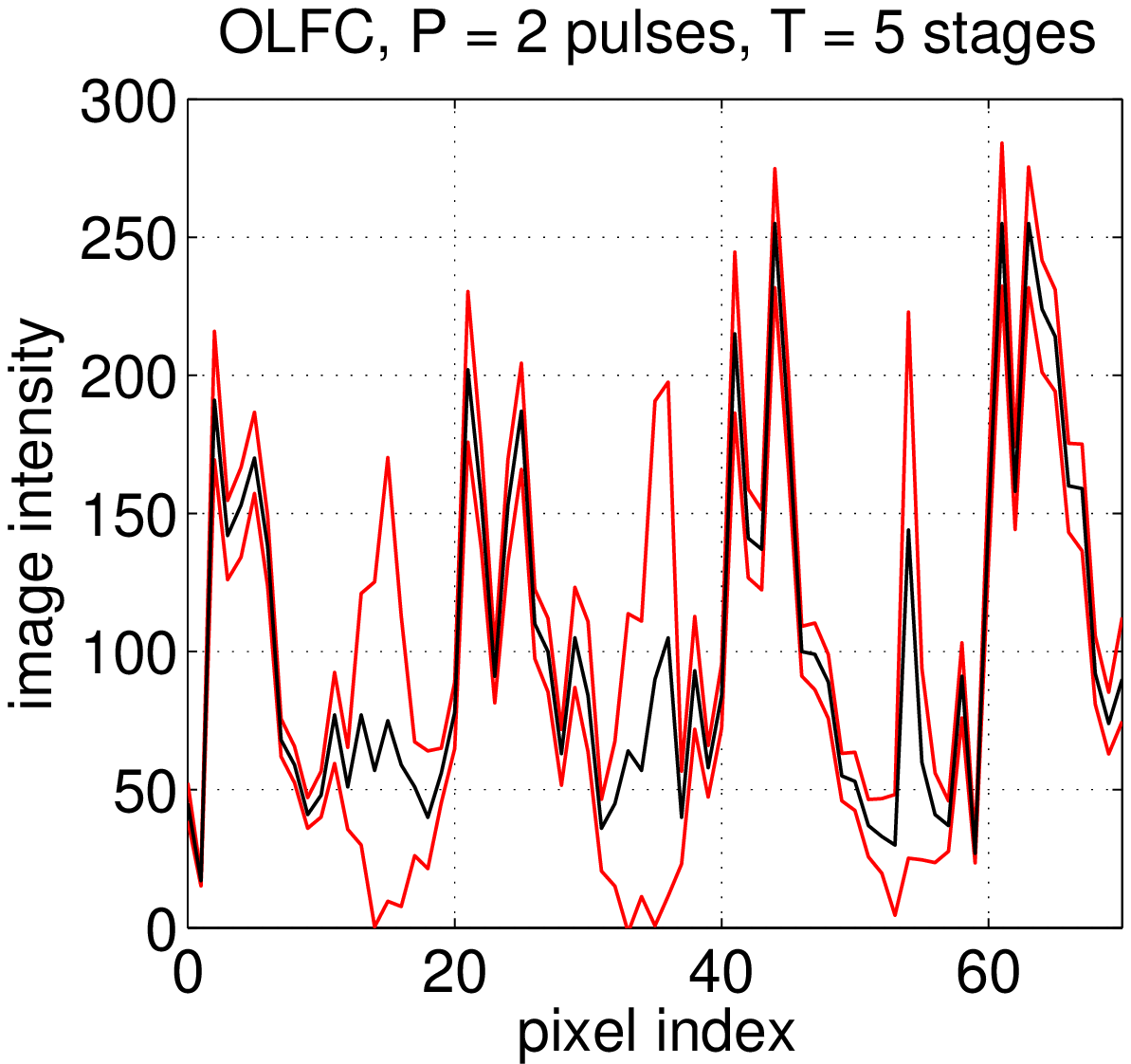}
\label{fig:profileOLFCT5}}
}
\caption{One-dimensional profiles passing through the line of tanks in Fig.~\ref{fig:tanksOrig}.  Middle curves indicate the true image intensities while upper and lower curves correspond to one standard deviation above and below the mean of $100$ reconstructions for each policy using $P = 2$ pulses per location.  The $5$-stage OLFC policy reduces the standard deviation by a further $2$--$3$ dB relative to the $2$-stage OLFC policy.}
\label{fig:profiles}
\end{figure}

In Fig.~\ref{fig:profiles}, we show one-dimensional profiles passing through the line of tanks.  The middle curves indicate the true image intensities while the upper and lower curves correspond to one standard deviation above and below the mean reconstruction for each policy, where the mean and standard deviation are computed from $100$ realizations.  The number of pulses per location is $P = 2$.  The variability in the reconstruction is clearly reduced using OLFC, in particular in the higher-amplitude regions corresponding to targets.  The $5$-stage OLFC policy further reduces the standard deviation by $2$--$3$ dB relative to the $2$-stage OLFC policy.

\section{Conclusions and future work}
\label{sec:concl}

We have presented multistage resource allocation policies for the sequential estimation of sparse signals under a variety of loss and effort functions.  Our formulation of the problem permits the application of techniques from dynamic programming, in particular open-loop feedback control.  The proposed policies improve monotonically with the number of stages and thus extend the optimal two-stage policy developed in \cite{bashan2008}.  Simulations and a radar imaging example also show gains relative to distilled sensing \cite{haupt2011} and dramatic improvements relative to non-adaptive sensing.

The dynamic programming approach taken in this paper is quite general and can potentially be leveraged to develop tractable policies for other inference tasks such as detection or a combination of detection and estimation.  More general observation models involving linear combinations may also be incorporated; the matched filtering in the radar example in Section \ref{sec:SARtanks} is only a preliminary step in this direction.  On the more theoretical side, the performance curves in Fig.~\ref{fig:gainSNR} motivate the need for bounds on the achievable performance of adaptive sensing that are more refined than the oracle bound.  Results in this vein for the case of a discrete resource budget have appeared recently 
\ifCLASSOPTIONdraftcls
\cite{ariascastro2013},\textcolor{red}{\cite{castro2012}}.
\else
\cite{ariascastro2013,castro2012}.
\fi

\appendices


\section{Proof of Lemma \ref{lem:probDist} and derivation of posterior probability distributions}
\label{app:probDist}

In this appendix, we prove Lemma \ref{lem:probDist} and indicate how the state variable recursions \eqref{eqn:stateEvol} are derived.
Attention is paid to the adaptive nature of the observations, specifically the dependence of the sensing effort $\mblambda(t)$ on past observations $\mbY(t)$.

First we show that the conditional distribution $f(\mbtheta \mid \mbI=\ones, \mbY(t))$ is independent Gaussian.
This can be done inductively starting with $t = 0$, in which case there are no observations and $f(\mbtheta \mid \mbI=\ones, \mbY(t))$ is given by the assumed independent Gaussian prior:
\begin{equation}\label{eqn:thetaPrior}
f(\mbtheta \mid \mbI=\ones) = \prod_{i=1}^{N} f(\theta_{i} \mid I_{i}=1) = \prod_{i=1}^{N} \phi(\theta_{i}; \mu_{i}(0), \sigma_{i}^{2}(0)).
\end{equation}
Next we assume that $f(\mbtheta \mid \mbI=\ones, \mbY(t-1))$ is given and use Bayes' rule to obtain the proportionality 
\ifCLASSOPTIONdraftcls
\begin{equation}\label{eqn:thetaPostRec}
f(\mbtheta \mid \mbI=\ones, \mbY(t)) \propto \\ f(\mby(t) \mid \mbtheta, \mbI=\ones, \mbY(t-1)) f(\mbtheta \mid \mbI=\ones, \mbY(t-1)) 
\end{equation}
\else
\begin{multline}\label{eqn:thetaPostRec}
f(\mbtheta \mid \mbI=\ones, \mbY(t)) \propto \\ 
f(\mby(t) \mid \mbtheta, \mbI=\ones, \mbY(t-1)) f(\mbtheta \mid \mbI=\ones, \mbY(t-1))
\end{multline}
\fi
as functions of $\mbtheta$.  Since conditioning on $\mbY(t-1)$ also fixes $\lambda_{i}(t-1)$ in \eqref{eqn:obs}, the observations $y_{i}(t)$ are conditionally independent and Gaussian and the likelihood term $f(\mby(t) \mid \mbtheta, \mbI=\ones, \mbY(t-1))$ simplifies to 
\begin{equation}\label{eqn:yGivenThetaI}
f(\mby(t) \mid \mbtheta, \mbI=\ones, \mbY(t-1)) = \prod_{i=1}^{N} \phi(y_{i}(t); \theta_{i}, \sigma^{2} / h(\lambda_{i}(t-1))).
\end{equation}
From \eqref{eqn:thetaPrior}--\eqref{eqn:yGivenThetaI} it can be seen that $\mbtheta \mid \mbI=\ones, \mbY(t)$ retains an independent Gaussian distribution for all $t$ with marginals given by 
\ifCLASSOPTIONdraftcls
\begin{equation}\label{eqn:thetaiPostRec}
f(\theta_{i} \mid I_{i}=1, \mbY(t)) \propto \phi(y_{i}(t); \theta_{i}, \sigma^{2} / h(\lambda_{i}(t-1))) f(\theta_{i} \mid I_{i}=1, \mbY(t-1)).
\end{equation}
\else
\begin{multline}\label{eqn:thetaiPostRec}
f(\theta_{i} \mid I_{i}=1, \mbY(t)) \propto \\ 
\phi(y_{i}(t); \theta_{i}, \sigma^{2} / h(\lambda_{i}(t-1))) f(\theta_{i} \mid I_{i}=1, \mbY(t-1)).
\end{multline}
\fi
%
We parameterize $f(\theta_{i} \mid I_{i}=1, \mbY(t))$ by its mean $\mu_{i}(t)$ and variance $\sigma_{i}^{2}(t)$ as in the statement of Lemma \ref{lem:probDist}.  A straightforward calculation 
starting from \eqref{eqn:thetaiPostRec} 
leads to the recursions in \eqref{eqn:muEvol} and \eqref{eqn:sigmaEvol}.  Solving \eqref{eqn:sigmaEvol} for the final-stage variance yields \eqref{eqn:sigmaFinal}.

We now show that the conditional probability mass function $p(\mbI \mid \mbY(t))$ is independent Bernoulli, proceeding inductively as before.  The base case $t = 0$ corresponds to the prior distribution, assumed to be i.i.d.~Bernoulli:
\begin{equation}\label{eqn:Iprior}
p(\mbI) = \prod_{i=1}^{N} p(I_{i}) = \prod_{i=1}^{N} p_{i}(0)^{I_{i}} (1-p_{i}(0))^{1-I_{i}}.
\end{equation}
Next we relate $p(\mbI \mid \mbY(t))$ to $p(\mbI \mid \mbY(t-1))$ using Bayes' rule:
\begin{equation}\label{eqn:IpostRec}
p(\mbI \mid \mbY(t)) 
= \frac{f(\mby(t) \mid \mbI, \mbY(t-1)) p(\mbI \mid \mbY(t-1))}{\sum_{\mbI'} f(\mby(t) \mid \mbI', \mbY(t-1)) p(\mbI' \mid \mbY(t-1))}.
\end{equation}
As before, conditioning on $\mbY(t-1)$ fixes $\lambda_{i}(t-1)$ in \eqref{eqn:obs} and thus $\mby(t) \mid \mbI, \mbY(t-1)$ is a linear combination of the independent random vectors $\mbtheta \mid \mbI, \mbY(t-1)$ and $\mbn(t)$.  Consequently we obtain
\ifCLASSOPTIONdraftcls
\begin{equation}\label{eqn:yGivenI}
f(\mby(t) \mid \mbI, \mbY(t-1)) = 
\prod_{i=1}^{N} \phi(y_{i}(t); I_{i}\mu_{i}(t-1), I_{i} \sigma_{i}^{2}(t-1) + \sigma^{2} / h(\lambda_{i}(t-1))).
\end{equation}
\else
\begin{multline}\label{eqn:yGivenI}
f(\mby(t) \mid \mbI, \mbY(t-1)) = \\
\prod_{i=1}^{N} \phi(y_{i}(t); I_{i}\mu_{i}(t-1), I_{i} \sigma_{i}^{2}(t-1) + \sigma^{2} / h(\lambda_{i}(t-1))).
\end{multline}
\fi
recalling that $f(\theta_{i} \mid I_{i}=1, \mbY(t-1))$ is parameterized by $\mu_{i}(t-1)$ and $\sigma_{i}^{2}(t-1)$.  From \eqref{eqn:Iprior}--\eqref{eqn:yGivenI} it can be concluded that the components of $\mbI \mid \mbY(t)$ remain independent with marginal distributions
\begin{equation}\label{eqn:IiPostRec}
p(I_{i} \mid \mbY(t)) 
= \frac{f(y_{i}(t) \mid I_{i}, \mbY(t-1)) p(I_{i} \mid \mbY(t-1))}{\sum_{I_{i}'=0}^{1} f(y_{i}(t) \mid I_{i}', \mbY(t-1)) p(I_{i}' \mid \mbY(t-1))}.
\end{equation}
The recursion for $p_{i}(t) = \Pr(I_{i}=1 \mid \mbY(t))$ in \eqref{eqn:pEvol} follows from \eqref{eqn:yGivenI} and \eqref{eqn:IiPostRec}.  
%

\section{Proof of Lemma \ref{lem:CME}}
\label{app:CME}

We first prove the lemma for loss functions of the form 
\begin{equation}\label{eqn:Ldelta}
L_{\delta}(a) = \begin{cases}
0, & 0 \leq a < \delta,\\
1, & a > \delta
\end{cases}
\end{equation}
for $\delta > 0$.  The expected loss for an estimate $\thetah$ is then 
\begin{equation}\label{eqn:expLossDelta}
\E\left[ L_{\delta}\left( \bigl\lvert \thetah - \theta \bigr\rvert \right) \right] = 1 - \int_{\thetah-\delta}^{\thetah+\delta} f(\theta) \, d\theta.
\end{equation}
By the symmetry and unimodality of $f(\theta)$ about $\mu$, it is intuitively clear and is formally proven in \cite{anderson1955} that the expected loss \eqref{eqn:expLossDelta} is minimized for $\thetah = \mu$.

A general non-decreasing loss function $L$ can be approximated arbitrarily closely by a sum of functions of the form in \eqref{eqn:Ldelta} in a manner reminiscent of Lebesgue integration.  Given a step size $\Delta L > 0$, we construct the approximation 
\[
\hat{L}(a) = \Delta L \sum_{k=1}^{\infty} L_{L^{-1}(k\Delta L)}(a),
\]
where $L^{-1}(k\Delta L)$ denotes the smallest value of $a$ such that $L(a) = k\Delta L$.  By the linearity of expectations, the expected value of $\hat{L}\left(\abs{\thetah - \theta}\right)$ is a sum of functions of the form in \eqref{eqn:expLossDelta}.  Since $\thetah = \mu$ minimizes each term in the sum individually, it also minimizes the overall sum and hence the mean estimate is optimal for $\hat{L}$.  As $\Delta L \to 0$, $\hat{L}$ converges to $L$ and the statement is proven for $L$.

\section{Solution of problem \texorpdfstring{\eqref{eqn:probOLFCMSE}}{(20)}}
\label{app:OLC}

%
%

For notational simplicity, we write $p_{i}$, $r_{i}$, $\lambda_{i}$, and $\Lambda$ in this appendix for the quantities $p_{i}(t)$, $\sigma^{2} / \sigma_{i}^{2}(t)$, $\lambdab_{i}(t)$, and $\Lambda(t)$ in \eqref{eqn:probOLFCMSE}.  We also use $J$ to denote the cost function.  As noted in Section \ref{subsec:policyOLFC}, \eqref{eqn:probOLFCMSE} is a convex minimization problem subject to a simplex constraint and therefore satisfies an optimality condition similar to \eqref{eqn:optCondSimplexMax}:
\begin{equation}\label{eqn:optCondSimplex}
\text{if} \quad \lambda_{i}^{\ast} > 0 \quad 
\text{then} \quad \frac{\partial J}{\partial\lambda_{i}}(\mblambda^{\ast}) \leq \frac{\partial J}{\partial\lambda_{j}}(\mblambda^{\ast}) \quad \forall \; j \neq i.
\end{equation}
Condition \eqref{eqn:optCondSimplex} implies that the optimal solution to \eqref{eqn:probOLFCMSE} satisfies an index rule in the sense that the non-zero components of the optimal solution correspond to the largest $p_{i}^{\gamma} / r_{i}$, where $\gamma = 2/(q+2)$.  
%
%
%
To prove this fact, suppose that $i$ and $j$ are such that $\lambda_{i}^{\ast} > 0$ and $\lambda_{j}^{\ast} = 0$ but $p_{i}^{\gamma}/r_{i} \leq p_{j}^{\gamma}/r_{j}$.  Then 
\[
\frac{\partial J}{\partial\lambda_{i}} = -\frac{q}{2} \frac{p_{i}}{(r_{i} + \lambda_{i}^{\ast})^{1/\gamma}} > -\frac{q}{2} \frac{p_{i}}{r_{i}^{1/\gamma}} \geq -\frac{q}{2} \frac{p_{j}}{r_{j}^{1/\gamma}} = \frac{\partial J}{\partial\lambda_{j}},
\]
contradicting the optimality condition \eqref{eqn:optCondSimplex}.  The index rule can be stated in terms of the permutation $\pi$ defined in \eqref{eqn:orderingOLFC}, which in the notation of this appendix sorts the quantities $p_{i}^{\gamma} / r_{i}$ in non-increasing order.  Specifically, we have $\lambda_{\pi(i)}^{\ast} > 0$ for $i = 1, \ldots, k$ for some integer $k$, $\lambda_{\pi(i)}^{\ast} = 0$ for $i = k+1, \ldots, N$, and $p_{\pi(k)}^{\gamma}/r_{\pi(k)} > p_{\pi(k+1)}^{\gamma}/r_{\pi(k+1)}$ strictly.

The optimality condition \eqref{eqn:optCondSimplex} also implies that the partial derivatives corresponding to non-zero components of the optimal solution must be equal.  Hence 
\begin{equation}\label{eqn:optCondSimplexNonzeros}
-\frac{2}{q} \frac{\partial J}{\partial\lambda_{\pi(i)}} = \frac{p_{\pi(i)}}{(r_{\pi(i)} + \lambda_{\pi(i)}^{\ast})^{1/\gamma}} = C^{-1/\gamma}, \quad i = 1, \ldots, k,
\end{equation}
where $C$ is a constant to be determined.  A slight rearrangement of \eqref{eqn:optCondSimplexNonzeros} yields the expression in \eqref{eqn:lambdab*} for $i = 1,\ldots,k$.  The value of $C$ in \eqref{eqn:C} is obtained by summing \eqref{eqn:lambdab*} over $i = 1, \ldots, k$ and noting that $\sum_{i=1}^{k} \lambda_{\pi(i)}^{\ast} = \sum_{i=1}^{N} \lambda_{i}^{\ast} = \Lambda$.
%
%
%
%
%
%

It remains to determine the cutoff index $k$.  This can be done by enforcing the condition $\lambda_{\pi(i)}^{\ast} > 0$ for $i = 1, \ldots, k$ and the optimality condition \eqref{eqn:optCondSimplex} for $j = \pi(k+1),\ldots,\pi(N)$ (corresponding to the zero-valued components).  The first condition is equivalent to 
\[
C > \frac{r_{\pi(i)}}{p_{\pi(i)}^{\gamma}}, \quad i = 1, \ldots, k,
\]
while the second is equivalent to 
\begin{equation}\label{eqn:k0}
C \leq \frac{r_{\pi(i)}}{p_{\pi(i)}^{\gamma}}, \quad i = k+1, \ldots, N.
\end{equation}
%
Given the definition of $\pi$ in \eqref{eqn:orderingOLFC}, the most stringent conditions correspond to $i = k$ and $i = k+1$, i.e., 
\begin{equation}\label{eqn:k1}
\frac{r_{\pi(k)}}{p_{\pi(k)}^{\gamma}} < \frac{\Lambda + \sum_{i=1}^{k} r_{\pi(i)}}{\sum_{i=1}^{k} p_{\pi(i)}^{\gamma}} \leq \frac{r_{\pi(k+1)}}{p_{\pi(k+1)}^{\gamma}},
\end{equation}
upon substituting \eqref{eqn:C}.  Solving \eqref{eqn:k1} for $\Lambda$ yields 
%
%
%
%
the condition $b(k-1) < \Lambda \leq b(k)$ using the definition of $b(k)$ in \eqref{eqn:bOLFC}.  If $k = N$, \eqref{eqn:k0} is absent and we only have the condition $\Lambda > b(N-1)$, or equivalently we may define $b(N) = \infty$.  Thus the number of non-zero components $k$ is determined by the interval $(b(k-1), b(k)]$ to which $\Lambda$ belongs.  This mapping from $\Lambda$ to $k$ is well-defined if $b(k)$ is a non-decreasing function of $k$ so that the intervals $(b(k-1), b(k)]$ are non-overlapping and span the positive real line.  Indeed we have  
\begin{align*}
b(k) &= \frac{r_{\pi(k+1)}}{p_{\pi(k+1)}^{\gamma}} \sum_{i=1}^{k} p_{\pi(i)}^{\gamma} - \sum_{i=1}^{k} r_{\pi(i)}\\
&\geq \frac{r_{\pi(k)}}{p_{\pi(k)}^{\gamma}} \sum_{i=1}^{k} p_{\pi(i)}^{\gamma} - \sum_{i=1}^{k} r_{\pi(i)}\\
&= \frac{r_{\pi(k)}}{p_{\pi(k)}^{\gamma}} \sum_{i=1}^{k-1} p_{\pi(i)}^{\gamma} - \sum_{i=1}^{k-1} r_{\pi(i)}\\
&= b(k-1),
\end{align*}
where the inequality is due to \eqref{eqn:orderingOLFC}.  



\ifCLASSOPTIONcaptionsoff
  \newpage
\fi



\bibliographystyle{IEEEtran}
\bibliography{IEEEabrv,adapSens,sparseLinInv,optimization}
\end{document}